\documentclass{aa}

\usepackage{lscape}
\usepackage{natbib}
\usepackage{graphicx}

\begin{document}

\title{The gas distribution in the outer regions of galaxy clusters}
\author{D. Eckert\inst{1,2} \and F. Vazza\inst{3}  \and S. Ettori\inst{4,5} \and S. Molendi\inst{1} \and D. Nagai\inst{6} \and E. T. Lau\inst{6,7} \and M. Roncarelli\inst{8} \and M. Rossetti\inst{1,9}  \and S. L. Snowden\inst{10} \and F. Gastaldello\inst{1,11}}
\institute{INAF - IASF-Milano, Via E. Bassini 15, 20133 Milano, Italy
\and
ISDC Data Centre for Astrophysics, Geneva Observatory, ch. d'Ecogia 16, 1290 Versoix, Switzerland\\
\email{Dominique.Eckert@unige.ch}
\and
Jacobs University Bremen, Campus Ring 1, 28759 Bremen, Germany
\and
INAF - Osservatorio Astronomico di Bologna, Via Ranzani 1, 40127 Bologna, Italy
\and
INFN, Sezione di Bologna, viale Berti Pichat 6/2, 40127 Bologna, Italy
\and
Department of Physics, Yale University, New Haven, CT 06520, USA
\and
Shanghai Astronomical Observatory, 80 Nandan Road, Shanghai 200030, China
\and
Dipartimento di Astronomia, Universit\`a di Bologna, via Ranzani 1, 40127 Bologna, Italy
\and
Dipartimento di Fisica, Universit\`a degli studi di Milano, via Celoria 16, 20133 Milano, Italy
\and
NASA/Goddard Space Flight Center, Code 662, Greenbelt, MD 20771, USA
\and
University of California at Irvine, 4129, Frederick Reines Hall, Irvine, CA, 92697-4575, USA
}
\abstract{}{We present our analysis of a local ($z=0.04-0.2$) sample of 31 galaxy clusters with the aim of measuring the density of the X-ray emitting gas in cluster outskirts. We compare our results with numerical simulations to set constraints on the azimuthal symmetry and gas clumping in the outer regions of galaxy clusters.}{We have exploited the large field-of-view and low instrumental background of \emph{ROSAT}/PSPC to trace the density of the intracluster gas out to the virial radius. We stacked the density profiles to detect a signal beyond $r_{200}$ and measured the typical density and scatter in cluster outskirts. We also computed the azimuthal scatter of the profiles with respect to the mean value to look for deviations from spherical symmetry. Finally, we compared our average density and scatter profiles with the results of numerical simulations.}{As opposed to some recent \emph{Suzaku} results, and confirming previous evidence from \emph{ROSAT} and \emph{Chandra}, we observe a steepening of the density profiles beyond $\sim r_{500}$. Comparing our density profiles with simulations, we find that non-radiative runs predict density profiles that are too steep, whereas runs including additional physics and/or treating gas clumping agree better with the observed gas distribution. We report high-confidence detection of a systematic difference between cool-core and non cool-core clusters beyond $\sim0.3r_{200}$, which we explain by a different distribution of the gas in the two classes. Beyond $\sim r_{500}$, galaxy clusters deviate significantly from spherical symmetry, with only small differences between relaxed and disturbed systems. We find good agreement between the observed and predicted scatter profiles, but only when the 1\% densest clumps are filtered out in the \texttt{ENZO} simulations.}{Comparing our results with numerical simulations, we find that non-radiative simulations fail to reproduce the gas distribution, even well outside cluster cores. Although their general behavior agrees more closely with the observations, simulations including cooling and star formation convert a large amount of gas into stars, which results in a low gas fraction with respect to the observations. Consequently, a detailed treatment of gas cooling, star formation, AGN feedback, and consideration of gas clumping is required to construct realistic models of the outer regions of clusters.}
\maketitle

\section{Introduction}

The outskirts of galaxy clusters are the regions where the transition between the virialized gas of clusters and the accreting matter from large-scale structure occurs and where the current activity of structure formation takes place. Around the virial radius, the assumption of hydrostatic equilibrium, which is a necessary assumption for reconstructing cluster masses from X-ray measurements, might not be valid any more \citep[e.g.,][]{evrard96}, which could introduce biases to X-ray mass proxies \citep{rasia,piffaretti,nagai07,lau09,meneghetti,fabjan}. As a result, the characterization of the X-ray emitting gas in the outer regions of galaxy clusters is important for mapping the gas throughout the entire cluster volume, studying the formation processes currently at work in the Universe, and performing accurate mass estimates for cosmological purposes \citep[e.g.,][]{allen11}.

Because of the low surface brightness of the X-ray emitting gas and the extended nature of the sources, measuring the state of the intracluster gas around the virial radius is challenging \citep{ettoriwfxt}. Recently, the \emph{Suzaku} satellite has achieved a breakthrough in this domain, performing measurements of cluster temperatures out to $r_{200}$\footnote{We define $r_\Delta$ as the radius within which $M(<r_\Delta)/\frac{4}{3}\pi r_\Delta^3=\Delta\rho_{crit}$} \citep{reip09,bautz,kawa,hoshino,simionescu,akamatsu,humphrey}, even in one case beyond $r_{200}$ \citep{george}, although this detection is likely hampered by systematic effects \citep{eckertpks}. Interestingly, some of the \emph{Suzaku} results indicate very steep temperature profiles and shallow density profiles in cluster outskirts, at variance with the results from \emph{XMM-Newton} \citep{pratt07,lm08,xmmcat,croston}, \emph{Chandra} \citep{vikhlinin06,ettbal}, \emph{ROSAT} \citep{vikhlinin99,neumann05}, and with the results from numerical simulations \citep{roncarelli,tozzi,nagai}. Thus, the behavior of the gas in cluster outskirts is still the subject of debate. Throughout paper, we refer to cluster outskirts as the region with $r>r_{500}$.

Thanks to its large field of view (FOV, $\sim 2$ deg$^2$) and low instrumental background, \emph{ROSAT}/PSPC is to the present day the most sensitive instrument for low surface-brightness emission. Its ability to detect cluster emission at large radii has been demonstrated by \citet{vikhlinin99} and \citet{neumann05} (hereafter, V99 and N05). Because of the large FOV, it can perform simultaneous local background measurements, so it is less affected than \emph{Suzaku} by systematic uncertainties. Its main limitation, however, is the restricted band pass and poor spectral resolution, which makes it impossible to measure cluster temperatures. 

This paper presents the analysis of a sample of 31 galaxy clusters observed with \emph{ROSAT}/PSPC, with the aim of characterizing the cluster emission at large radii and comparing the results with three different sets of numerical simulations \citep{roncarelli,nagai,vazza10}.  The paper is organized as follows. In Sect. \ref{secsample}, we describe our cluster sample and the available data. We present our data analysis technique in Sect. \ref{secdata} and report our results in Sect. \ref{secres}. We compare our results with numerical simulations in Sect. \ref{secsim} and discuss them in Sect. \ref{secdisc}.

Throughout the paper, we assume a $\Lambda$CDM cosmology with $\Omega_m=0.3$, $\Omega_\Lambda=0.7$, $\Omega_b=0.047$, and $H_0=70$ km s$^{-1}$ Mpc$^{-1}$.

\section{The sample}
\label{secsample}

We selected objects in the redshift range $0.04-0.2$, such that $r_{200}$ is easily contained within the FOV of the instrument and is large enough to allow for an adequate sampling of the density profile. We restricted ourselves to observations with enough statistics to constrain the emission around the virial radius. Our final sample comprises 31 clusters in the temperature range 2.5-9 keV, with the addition of A2163 ($kT\sim18$ keV). Among our sample, we classified 14 clusters as cool core (CC) following the classification of \citet{cavagnolo} (i.e. they exhibit a central entropy $K_0<30$ keV cm$^2$), and 17 as non cool core (NCC, $K_0>30$ keV cm$^2$). We recall that CC clusters exhibit a relaxed morphology, a high central density and a temperature decrement in the central regions, while NCCs trace dynamically-disturbed clusters with irregular morphologies and flat temperature and density profiles in their cores \citep[e.g.,][]{sanderson,hudson}.

\begin{figure*}
\resizebox{\hsize}{!}{\hbox{\includegraphics{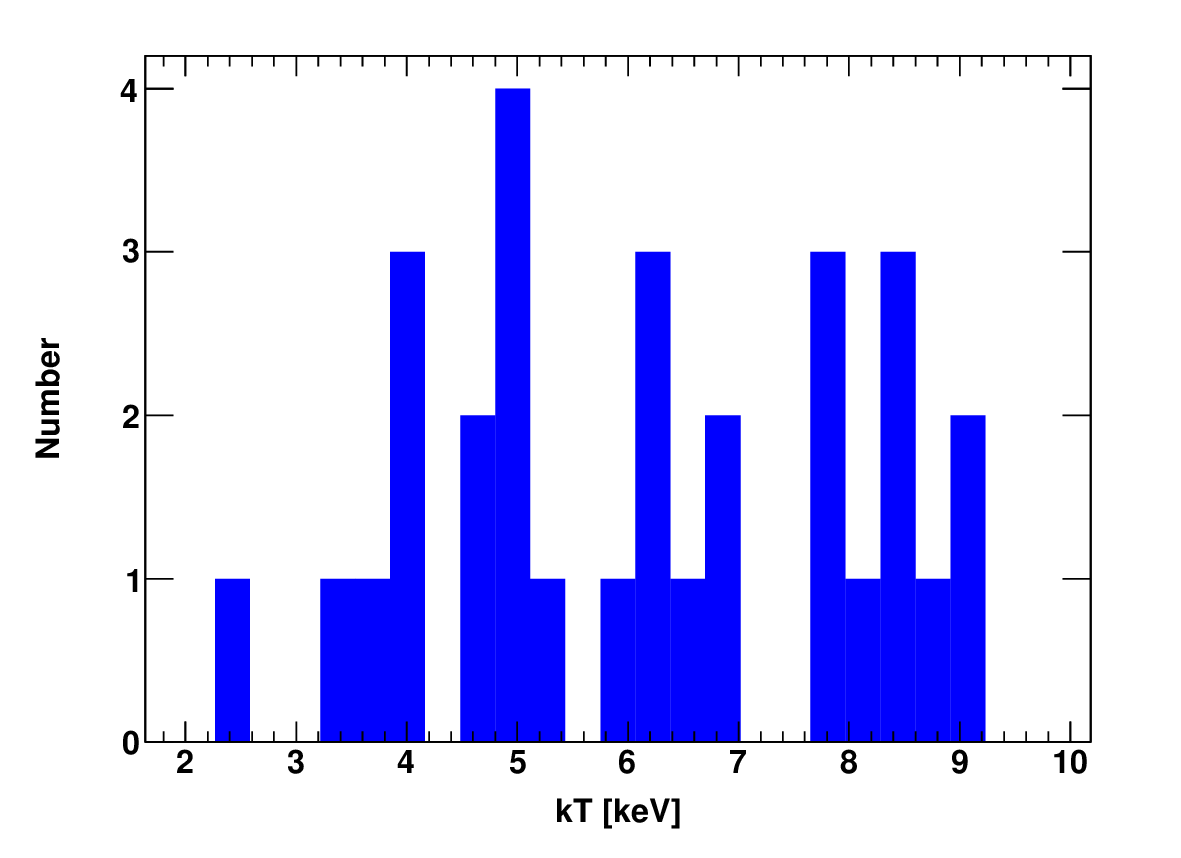}\includegraphics{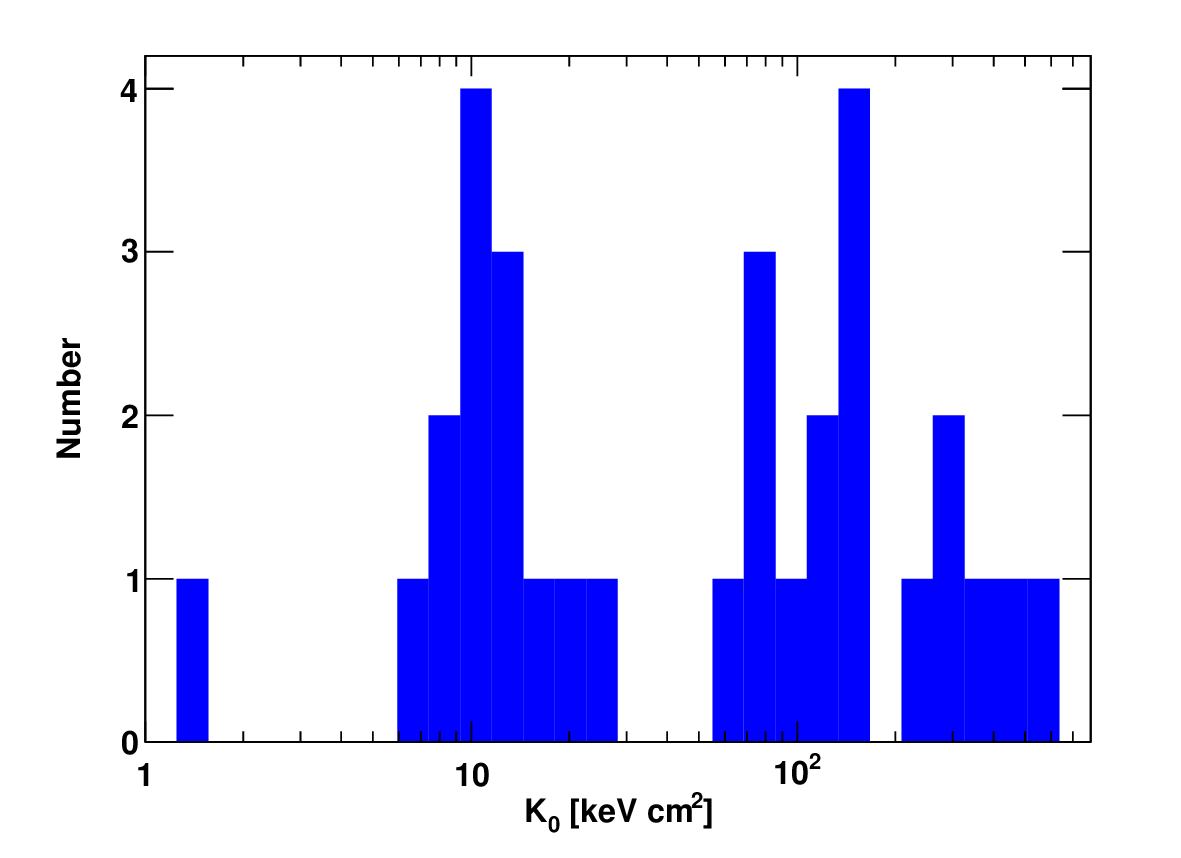}}}
\caption{Distribution of temperature (left) and central entropy (right) of the members of our sample (see Table \ref{master}). In the left panel, A2163 ($kT\sim18$ keV) is located outside of the range.}
\label{distribution}
\end{figure*}

Our sample of clusters, together with the log of the available data and some important quantities, is shown in Table \ref{master}. In Fig. \ref{distribution} we plot the distribution of temperature (left hand panel) and central entropy (right hand panel) for our sample. It should be noted that the sample was selected based on the quality of the existing observations and might be subject to selection effects. However, for the purpose of this work we did not require that the sample be representative or complete, since we are interested in characterizing cluster outskirts, which exhibit a high level of self-similarity.

\section{Data analysis}
\label{secdata}

\subsection{Data reduction}

We used the \emph{ROSAT} Extended Source Analysis Software \citep{esas} for data reduction. We filtered out time periods when the master veto count rate exceeds 220 cts/sec (using \texttt{valid\_times}), and extracted light curves for the whole observation using \texttt{rate\_pspc}. We used the \texttt{ao} executable to model the atmospheric column density for the scattering of solar X-rays, and fit the light curves in each energy band to get the relative contributions of the scattered solar X-rays (SSX) and of the long-term enhancements (LTE), using the \texttt{rate\_fit} executable.

We then extracted event images in each energy band and the corresponding effective exposure maps, taking vignetting effects into account. We computed the contribution of the various background components, the LTE (\texttt{lte\_pspc}), the particle background (\texttt{cast\_part}), and the SSX (\texttt{cast\_ssx}), and combined them to get a map of all the non-cosmic background components.

\subsection{Surface-brightness profiles}

The point-spread function (PSF) of \emph{ROSAT}/PSPC strongly depends on angle, and ranges from $\sim15$ arcsec on-axis to 2 arcmin in the outer parts of the FOV. Thus, the sensitivity of the instrument to point sources is higher on-axis, and a larger fraction of the cosmic X-ray background (CXB) is resolved. Consequently, when detecting sources in the image it is important to use a constant flux threshold, such that the same fraction of the CXB is resolved over the entire FOV and the value measured in the source-free regions can be used to subtract the background. We detect point sources using the program \texttt{detect} with a minimum count rate of 0.003 cts/sec in the R3-7 band ($\sim3\times10^{-14}$ ergs cm$^{-2}$ s$^{-1}$ in the 0.5-2.0 keV band) to resolve the same fraction of the CXB over the FOV, and mask the corresponding areas. To compute surface-brightness profiles, we extract count profiles from the event images in the R3-7 band (0.42-2.01 keV) with 30 arcsec bins centered on the surface-brightness peak, out to the radius of 50 arcmin. We divide each pixel by its corresponding exposure to account for the vignetting effects, following the procedure of \citet{eckert}\footnote{http://www.iasf-milano.inaf.it/$\sim$eckert/newsite/Proffit.html  }. We perform the same operation for the background map and subtract the non-cosmic background profile in each bin.

\begin{figure}
\resizebox{\hsize}{!}{\includegraphics{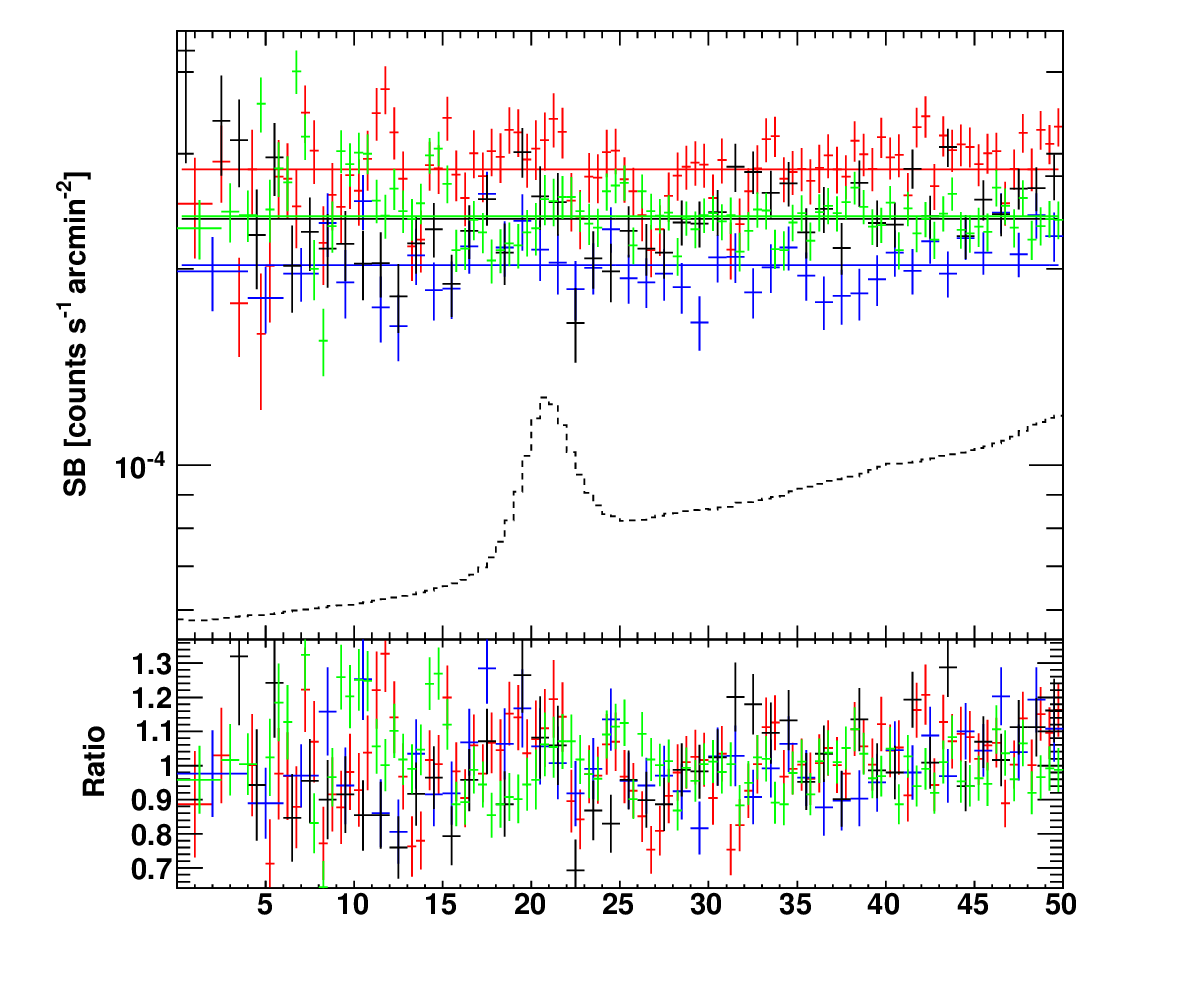}}
\caption{Surface-brightness profiles for 4 blank-field PSPC observations from the center of the FOV, fitted with a constant. The dashed line shows the vignetting correction curve for comparison, in arbitrary units; the bump at $\sim22$ arcmin is caused by the support structure. The bottom panel shows the ratio between data and model.}
\label{lockman}
\end{figure}

We tested this procedure on four different blank fields to estimate the accuracy in our determination of the CXB. We extracted the surface-brightness profile for the four observations from the center of the FOV, grouped the bins to ensure a minimum of 100 counts per bin, and fitted the resulting profiles with a constant (see Fig. \ref{lockman}). While the agreement is qualitatively good, significant deviations from the model are found, leading to an excess scatter of $\sim 6\%$, which we used as an estimate of the systematic uncertainties in measuring the CXB. This value encompasses both the cosmic variance and the true systematic uncertainties, e.g., in the vignetting correction or determination of the particle background. The higher level of scatter in the central regions is explained by the small area of the corresponding annuli, which implies a large cosmic variance likely due to discrete sources with fluxes just under our exclusion threshold. Since in most cases the value of $r_{200}$ is larger than 15 arcmin, our systematic error of 6\% is a conservative estimate of the level of systematic uncertainties at the virial radius.

For each cluster, we then use temperature profiles from the literature (\emph{XMM-Newton}, \citet{xmmcat}; \emph{Chandra}, \citet{cavagnolo}; \emph{BeppoSAX}, \citet{sabrina}) to estimate the virial temperature of the cluster. We approximated $T_{vir}$ as the mean temperature in the 200-500 kpc region, i.e. excluding the cool core and the temperature decline in the outskirts \citep{lm08}. Using this estimate of $T_{vir}$, we computed the value of $r_{200}$ from the scaling relations of \citet{arnaud}. One might argue that the scaling relations of \citet{arnaud} were computed using the mean temperature in the $0.1-0.5r_{200}$ region, which in most cases extends beyond the available temperature profiles. Using the mean temperature profiles of \citet{lm08}, we computed the mean temperature extracted in the $0.1-0.5r_{200}$ and 200-500 kpc regions. In the temperature range of our sample, we found that the results differ at most by 2\%, so our values of $r_{200}$ are unbiased. We then used the source-free region of the observation ($r>1.3 r_{200}$) to fit the surface-brightness profile with a constant and get the cosmic background level for the observation, with the exception of the Triangulum Australis cluster, for which we used the range $r>1.1r_{200}$ because of the high value of $r_{200}$ ($\sim37$ arcmin).

After having estimated the sky background for our observation, we again extracted the surface-brightness profile in the radial range $0-1.3 r_{200}$ with logarithmic bin size. The best-fit value for the CXB was subtracted from the profile and its error was added in quadrature to each bin. The systematic error of 6\% on the CXB was also added in quadrature to account for the cosmic variance and systematic uncertainties. For comparison, we note that in most cases the statistical uncertainties in the profiles are on the order of 10\% of the CXB value around $r_{200}$.

\subsection{Density profiles}

To compute the density profiles, we first rebinned our background-subtracted surface-brightness profiles to ensure a minimum of 200 counts per bin and a detection significance of at least $3\sigma$, to reach sufficient statistics in each bin. We used the procedure of \citet{kriss} to deproject the observed profiles, and the PSPC response to convert the observed count rates into emission measure, through the normalization of the MEKAL model \citep[see][for details]{eckertpks},
\begin{equation}Norm=\frac{10^{-14}}{4\pi [d_A(1+z)]^2} \int n_e n_H dV ,\label{meknorm}\end{equation}
which is proportional to the emission measure. We assumed that the spectrum of our sources is described by an absorbed MEKAL model with $N_H$ fixed to the 21cm value \citep{kalberla} and abundance fixed to $0.3Z_\odot$. We used temperature profiles from the literature (see Table \ref{master}) and interpolated them onto the same grid as the SB profiles. The resulting model was then folded with the PSPC response, and the conversion from PSPC R3-7 count rate to emission measure was inferred. Beyond the limit of the temperature profiles, the temperature of the outermost annulus was used. We note that the conversion from PSPC count rate to emission measure is highly insensitive to the temperature; between 2 and 8 keV, the conversion factor changes at most by 4\%. Once converted into the MEKAL normalization, we inferred the density profiles, assuming spherical symmetry and constant density into each shell.

The error bars on the density profiles were estimated using a Monte Carlo approach. In each case, we generated $10^4$ realizations of the surface-brightness profile using Poisson statistics, and performed the geometrical deprojection following the method described above. The $1\sigma$ error bars were then estimated by computing the root-mean square deviation (RMS) of our $10^4$ realizations of the density profile in each density bin.

\subsection{Azimuthal scatter profiles}
\label{azmet}

For the purpose of this work, we are also interested in the deviations in the X-ray emission from spherical symmetry. We divide our images into $N$ azimuthal sectors with constant opening angle, and compute the surface-brightness profiles in each sector individually. We then compute the scatter of the various sectors with respect to the mean profile, following the definition introduced by \citet{vazzascat},

\begin{equation} \Sigma^2=\frac{1}{N}\sum_{i=1}^N \frac{(SB_i-\langle SB\rangle)^2}{\langle SB\rangle^2}, \label{stot}\end{equation}

\noindent where $\langle SB\rangle$ is the mean surface-brightness and $SB_i, i=1..N$ denotes the surface-brightness computed in the various sectors. It must be noted that the statistical fluctuations of the SB between the different sectors introduce a certain level of scatter in Eq. \ref{stot}, which must be taken into account for determining the level of intrinsic scatter. We used two different methods to disentangle between statistical and intrinsic scatter. In the first case, we computed the level of statistical scatter independently and subtracted it from Eq. \ref{stot}. In the second case, we used a maximum-likelihood estimator to determine the intrinsic scatter and its uncertainties. The two methods gave consistent results and are described in detail in Appendix \ref{appscat}. For the remainder of the paper, we refer to the results obtained using the direct method (see Sect. \ref{metone}).

In our analysis, we group the bins of the total surface-brightness profiles to reach a minimum of $8\sigma$ per bin to ensure adequate statistics in the scatter measurements, and then divide our images into 12 sectors with an opening of 30$^\circ$. The result of this analysis is a radial profile describing the intrinsic azimuthal scatter of the X-ray surface brightness, in percent.

It must be noted that the method presented here is sensitive to all kinds of deviations from spherical symmetry, whether it is induced by the asymmetry of the large-scale structure (e.g., filaments), by gas clumping or by ellipticity. The cause of the observed asymmetry cannot be determined from the azimuthal scatter alone. 

\section{Results}
\label{secres}

\subsection{Emission measure and density profiles}

\begin{figure*}
\resizebox{\hsize}{!}{\hbox{\includegraphics{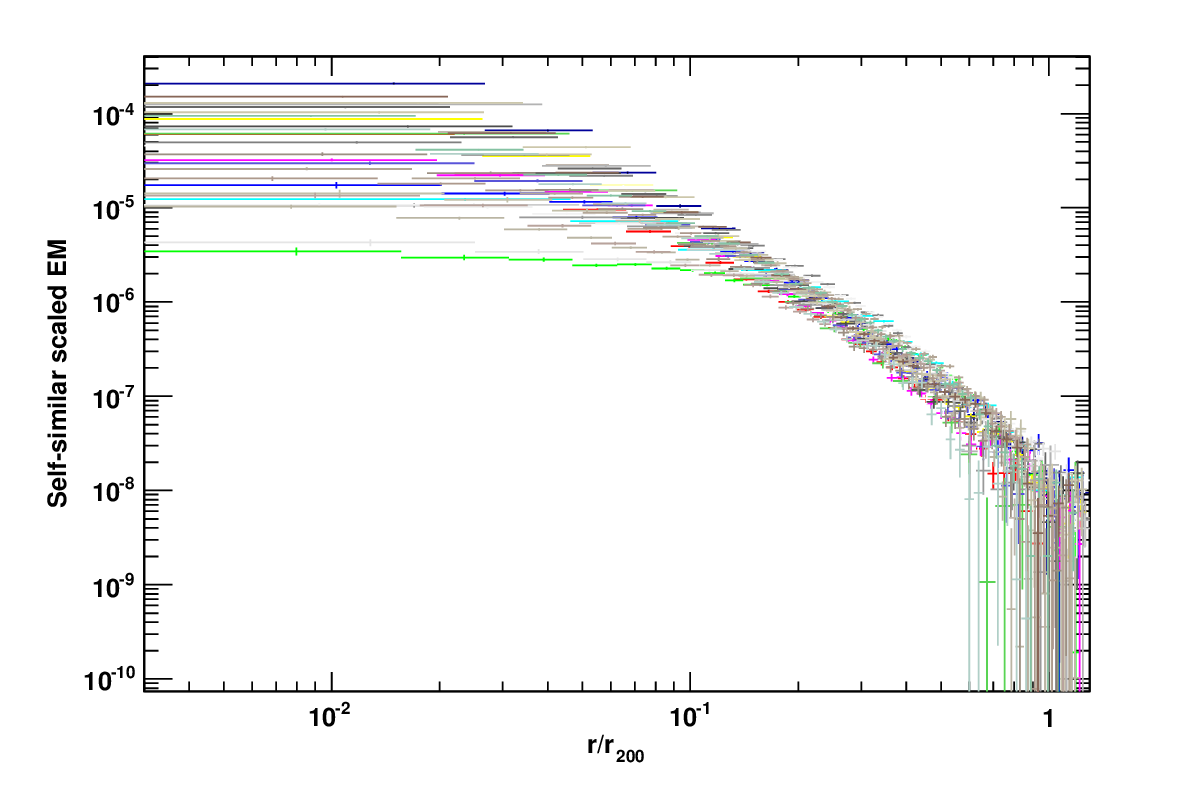}\includegraphics{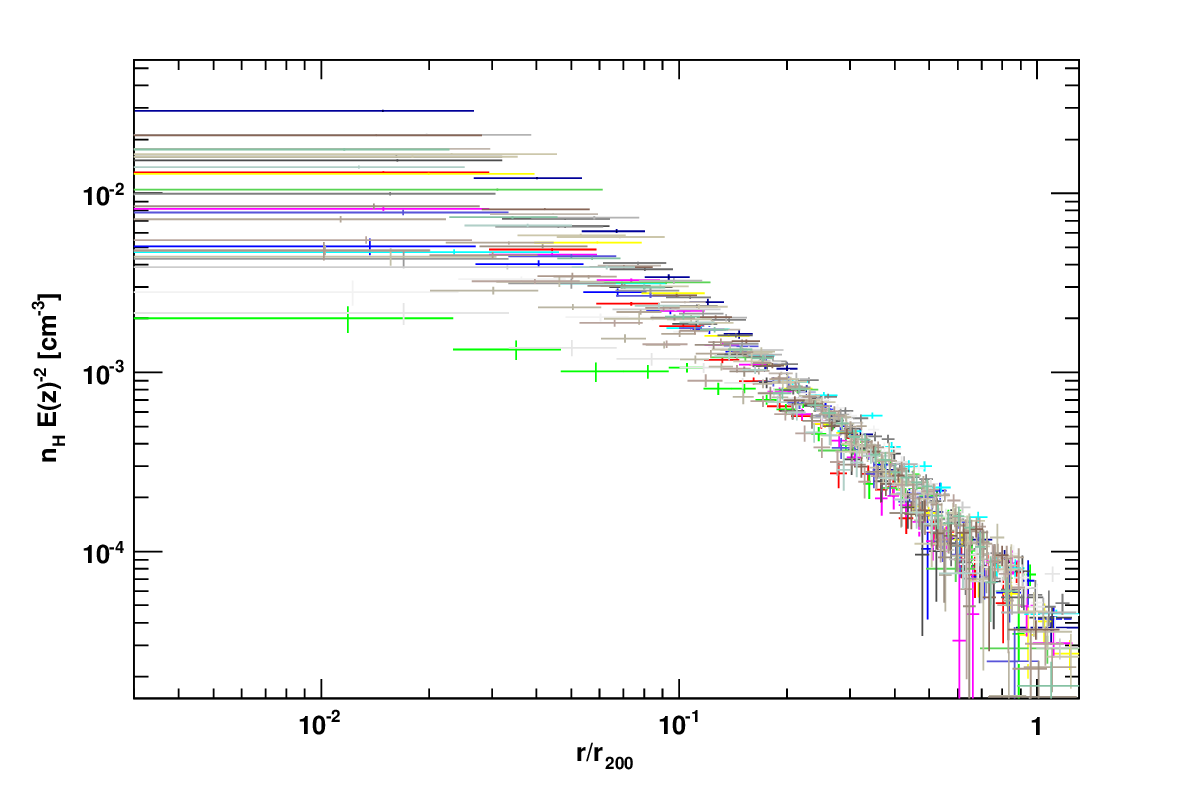}}}
\caption{Scaled emission measure (left, in units of cm$^{-6}$ Mpc) and density profiles (right) for the 31 clusters of our sample (see Table \ref{master}).}
\label{figall}
\end{figure*}

In Fig. \ref{figall} we show the scaled emission measure profiles (left, following Eq. \ref{meknorm}) and the deprojected density profiles (right) for the 31 clusters in our sample. A self-similar scaling was applied to the emission-measure profiles \citep{arnaud02}; i.e., each profile was rescaled by the quantity

\begin{equation}\Delta_{SSC}=\Delta_z^{2/3}(1+z)^{9/2}\left(\frac{kT}{10\mbox{ keV}}\right)^{1/2}.\end{equation}

\noindent The density profiles were rescaled by $E^2(z)=\Omega_m(1+z)^3+\Omega_\Lambda$ following their expected evolution with redshift \citep{croston}. As already noted by several authors \citep[e.g.,][]{vikhlinin99,neumann05,croston,lrm09}, the profiles show a remarkable level of self-similarity outside of the core ($r>0.2r_{200}$). On the other hand, the large scatter observed in the central regions reflects the distinction of the cluster population into CCs, showing a prominent surface-brightness peak, and NCCs, which exhibit a flat surface brightness profile in their cores, as expected from the standard $\beta$-model \citep{cavaliere},

\begin{equation}SB(r)=SB_0\left(1+\left(\frac{r}{r_c}\right)^2\right)^{-3\beta+0.5}.\label{beta}\end{equation}

\noindent In the radial range $0.2-0.7r_{200}$, the scatter of the density profiles is 10\%-20\%, in excellent agreement with the \emph{Chandra} \citep{vikhlinin06} and \emph{XMM-Newton} results \citep{croston}. However, \citet{croston} needed to rescale the profiles by $T^{-1/2}$ to account for the lower gas fraction in low-mass objects. In our case, performing such a scaling does not reduce the scatter of the profiles further. This is probably explained by the relatively narrow temperature range spanned in our sample (all but one objects have a temperature higher than 3 keV), such that the clusters in our sample should show little dependence on gas fraction.

\subsection{Stacked emission-measure profiles}
\label{secem}

To compute the mean profile of our sample, we interpolated each profile following a predefined binning in units of $r_{200}$ common to all clusters and performed a weighted mean to compute stacked profiles. The errors on the interpolated points were propagated to the stacked profiles. We also divided our sample into the two classes (CC and NCC) to look for differences between them.

\begin{figure}
\resizebox{\hsize}{!}{\includegraphics{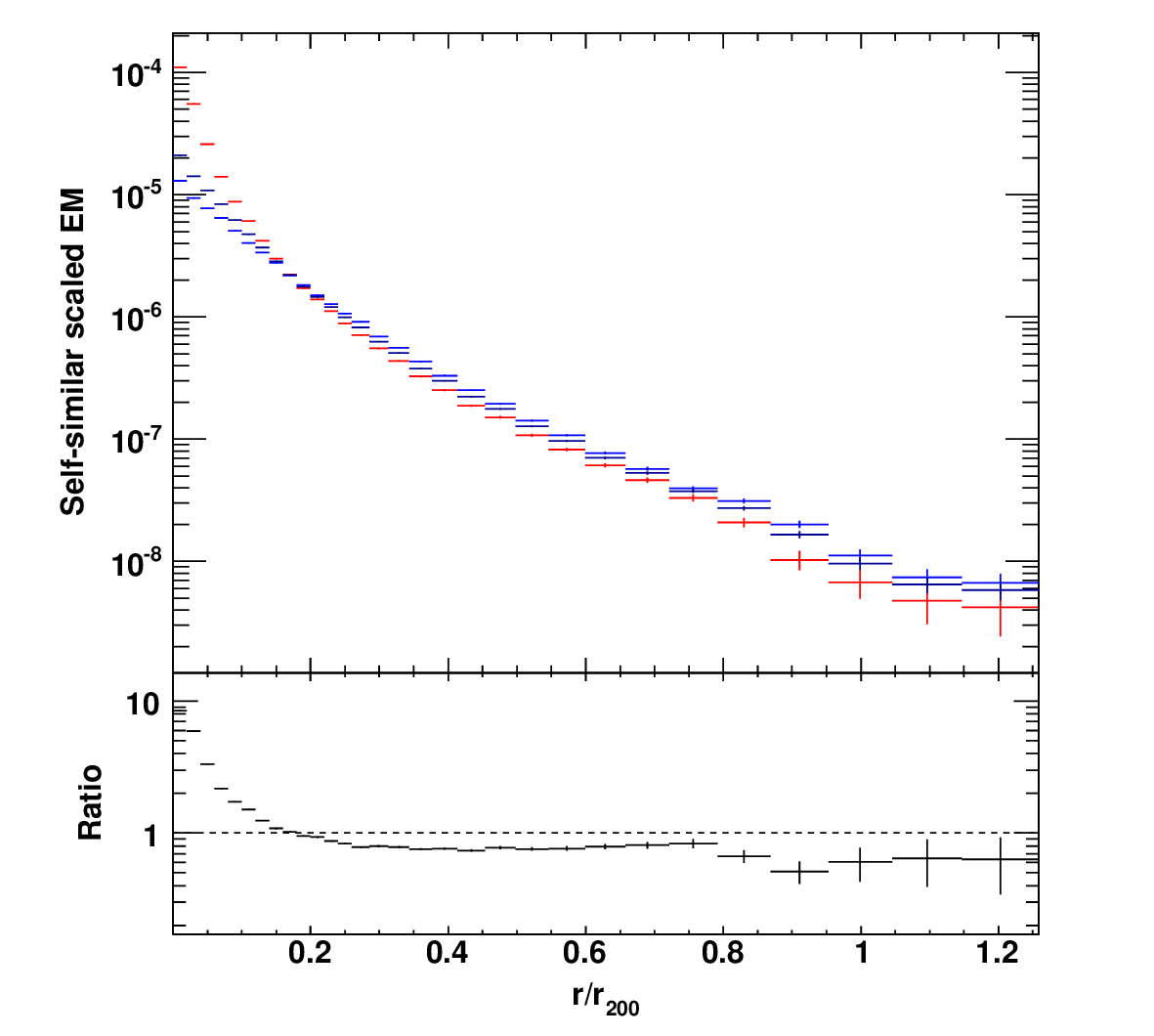}}
\caption{Stacked emission measure profile (in units of cm$^{-6}$ Mpc) for the entire sample (black), and the two populations individually (CC, red; NCC, blue). See also Appendix \ref{appem}. The bottom panel shows the ratio between the CC and NCC populations.}
\label{meanem}
\end{figure}

In Fig. \ref{meanem} we show the stacked emission-measure (EM) profile for the entire sample compared to the profiles stacked for the two populations separately (see also Appendix \ref{appem}). Interestingly, we note a clear distinction between the two classes in cluster outskirts (see the bottom panel of the figure). Namely, beyond $\sim0.3 r_{200}$, NCC profiles systematically exceed CCs. A similar effect has recently been noted by \citet{maughan11}, who found a crossing of the average density profiles at a similar radius, and also at a lower statistical significance in the works of \citet{arnaud10} and \citet{pratt10}. We stress that this effect is really a difference between the two classes; i.e. it is not introduced by a biased distribution of another quantity (such as temperature or redshift). Indeed, grouping the profiles according to the temperature or the redshift did not show any particular behavior, which indicates that we are really finding an intrinsic difference between the CC and NCC classes. This result could follow from a different distribution of the gas in the two populations or from a higher clumping factor in disturbed objects (see Sect. \ref{secdisc}).

Alternatively, the observed difference could be explained by an inaccurate determination of $r_{200}$ for NCC clusters. Indeed, the scaling relations of \citet{arnaud} were computed under the assumption of hydrostatic equilibrium, which is fulfilled better in CC clusters. This explanation is, however, unlikely. Indeed, to recover self-similarity, our value of $r_{200}$ should have been systematically \emph{underestimated} by $\sim 10\%$ for NCCs, i.e. since $r_{200}\propto T_{vir}^{1/2}$ the virial temperature of the NCC clusters should have been underestimated by more than 20\%. From mock \emph{Chandra} observations of a sample of simulated galaxy clusters, \citet{nagai07} have determined that the spectroscopic temperatures of unrelaxed clusters differs from that of relaxed clusters by $\sim5\%$, which is not enough to explain the observed difference. It is therefore unlikely that such a large error on the virial temperature would be made.

We fit the mean scaled emission-measure profiles from Fig. \ref{meanem} with the standard $\beta$-model (Eq. \ref{beta}), adding a second $\beta$ component in the case of the CC clusters to take the cool core into account. The (double) $\beta$ model gives a good representation of the data in the radial range $0-0.7r_{200}$ ($\sim r_{500}$), but significantly exceeds the observed profiles above this radius, in agreement with the results of V99, N05, and \citet{ettbal}. For CC clusters, the best-fit model gives $\beta=0.717\pm0.005$, while for NCC clusters we find $\beta=0.677\pm0.002$. Fitting the radial profiles in the range 0.65-1.3$r_{200}$, we observe a significant steepening, with a slope $\beta=0.963\pm0.054$ for CCs and $\beta=0.822\pm0.029$ for NCCs. As explained above, the slope of the NCC profile is flatter than that of the CC profile beyond $r_{500}$. The fits of the profiles in various radial ranges are reported in Table \ref{betavalues} to quantify the steepening.

Given the limited number of objects in our sample, we have to verify that this result is not a chance realization. We fit all the emission-measure profiles at $r>0.3r_{200}$ with a $\beta$ profile, fixing the value of $\beta$ to 0.7 and $r_c$ to $0.12r_{200}$, and extracted the best-fit normalization for all profiles. We then sorted the normalization values into the CC and NCC classes, and performed a Kolmogorov-Smirnov test to determine the probability that they originate in the same parent distribution. Using this procedure, we found that the chance probability for this result is very low, $P\sim6\times10^{-7}$. Therefore, we can conclude with good confidence that we are indeed finding an intrinsic difference between the two classes.

\subsection{Stacked density profiles}

\begin{figure}
\resizebox{\hsize}{!}{\includegraphics{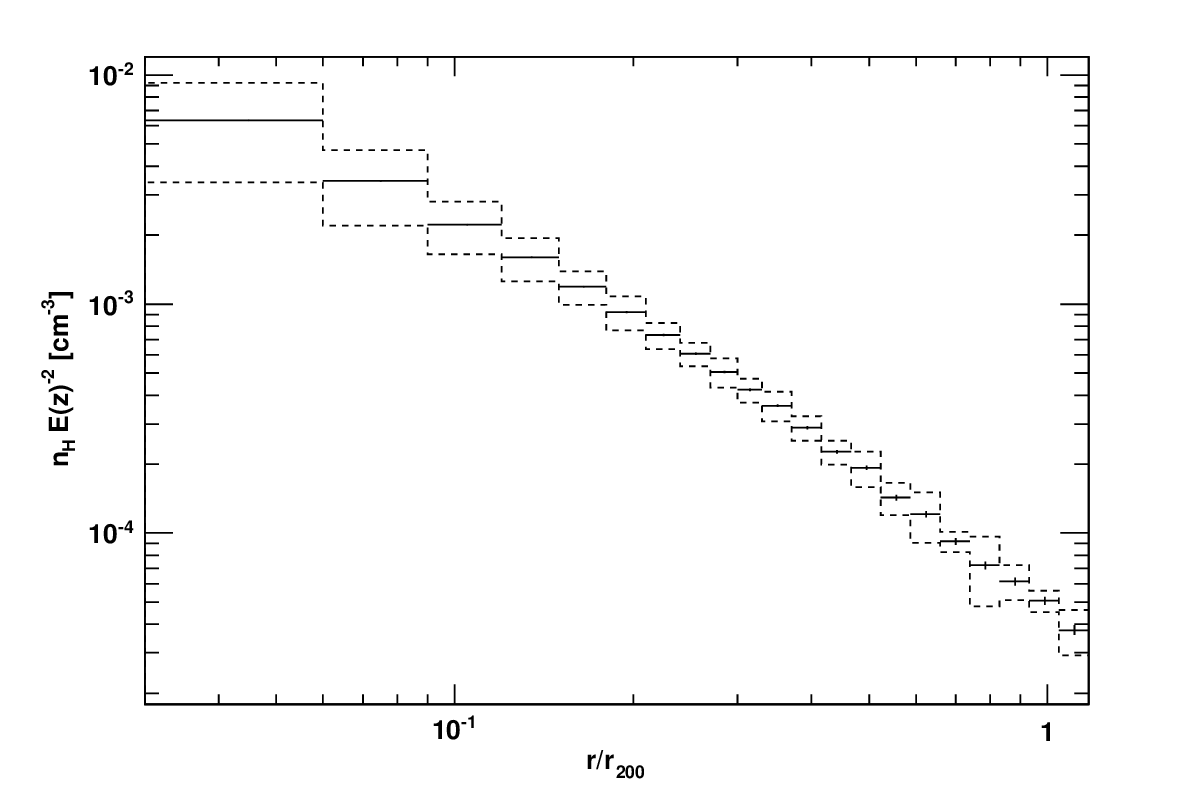}}
\caption{Average proton density profile for the entire sample. The dashed lines indicate the positive and negative scatter of the profiles around the mean value.}
\label{meandens}
\end{figure}

We stacked the density profiles shown in the right hand panel of Fig. \ref{figall} following the same method as for the EM profiles. From the different profiles, we also computed the scatter of the profiles around the mean value, following a method similar to the one presented in Sect. \ref{azmet} for the azimuthal scatter. The statistical scatter was subtracted from the total scatter using the same technique. In Fig. \ref{meandens} we show the average density profile of our clusters together with the scatter of the individual profiles around the mean value (see also Table \ref{tabmean}). At $r_{200}$, the mean density is $n_{200}=(3.8\pm0.4)\times10^{-5}\, E^2(z)$ cm$^{-3}$, with 25\% scatter. For comparison, it is interesting to note that the density of PKS 0745-191 claimed in the \emph{Suzaku} analysis of \citet{george} at $r_{200}$ deviates from our mean value by more than 5$\sigma$, which casts even more doubt on this measurement \citep{eckertpks}.

As for the EM, we also extracted mean density profiles individually for the two classes of clusters in our sample. The same behavior is observed at large radii; i.e., the density of NCC clusters is systematically higher (by $\sim$15\%) than that of CCs above $r\sim0.3 r_{200}$. A global steepening of the density profiles is also observed beyond $\sim r_{500}$.

Our density profiles are in good agreement with the results of V99. However, while V99 estimated the density from $\beta$-model fitting, we performed a geometrical deprojection of the data using temperature profiles to infer the mean density profile. This method has the advantage of not depending on any model.

\begin{center}
\begin{table}
\caption{\label{tabmean}Mean emission-measure and density profiles computed from our sample. }
\begin{tabular}{ccccc}
\hline
$R_{in}$ & $R_{out}$ & ScEM & $n_{H}E(z)^{-2}$ & $\sigma$ \\ 
\hline
\hline
0 & 0.03 & $(1.78 \pm 0.01)\cdot10^{-5}$ & $11.447 \pm 0.033 $ & 58 \\ 
0.03 & 0.06 & $(1.23 \pm 0.00)\cdot10^{-5}$ & $6.325 \pm 0.018 $ & 46 \\ 
0.06 & 0.09 & $(7.34 \pm 0.03)\cdot10^{-6}$ & $3.446 \pm 0.012 $ & 36 \\ 
0.09 & 0.12 & $(5.13 \pm 0.02)\cdot10^{-6}$ & $2.222 \pm 0.010 $ & 26 \\ 
0.12 & 0.15 & $(3.49 \pm 0.01)\cdot10^{-6}$ & $1.599 \pm 0.009 $ & 21 \\ 
0.15 & 0.18 & $(2.44 \pm 0.01)\cdot10^{-6}$ & $1.191 \pm 0.008 $ & 17 \\ 
0.18 & 0.21 & $(1.65 \pm 0.01)\cdot10^{-6}$ & $0.923 \pm 0.007 $ & 17 \\ 
0.21 & 0.24 & $(1.24 \pm 0.01)\cdot10^{-6}$ & $0.731 \pm 0.007 $ & 13 \\ 
0.24 & 0.27 & $(9.66 \pm 0.06)\cdot10^{-7}$ & $0.606 \pm 0.006 $ & 12 \\ 
0.27 & 0.30 & $(7.19 \pm 0.05)\cdot10^{-7}$ & $0.506 \pm 0.006 $ & 15 \\ 
0.30 & 0.33 & $(5.50 \pm 0.04)\cdot10^{-7}$ & $0.422 \pm 0.005 $ & 12 \\ 
0.33 & 0.37 & $(4.20 \pm 0.04)\cdot10^{-7}$ & $0.360 \pm 0.005 $ & 15 \\ 
0.37 & 0.42 & $(3.08 \pm 0.03)\cdot10^{-7}$ & $0.289 \pm 0.005 $ & 12 \\ 
0.42 & 0.47 & $(2.11 \pm 0.02)\cdot10^{-7}$ & $0.227 \pm 0.004 $ & 12 \\ 
0.47 & 0.52 & $(1.53 \pm 0.02)\cdot10^{-7}$ & $0.193 \pm 0.004 $ & 18 \\ 
0.52 & 0.59 & $(1.05 \pm 0.02)\cdot10^{-7}$ & $0.143 \pm 0.004 $ & 16 \\ 
0.59 & 0.66 & $(7.16 \pm 0.15)\cdot10^{-8}$ & $0.121 \pm 0.004 $ & 25 \\ 
0.66 & 0.74 & $(5.12 \pm 0.14)\cdot10^{-8}$ & $0.092 \pm 0.003 $ & 10 \\ 
0.74 & 0.83 & $(3.36 \pm 0.12)\cdot10^{-8}$ & $0.072 \pm 0.003 $ & 34 \\ 
0.83 & 0.93 & $(1.97 \pm 0.12)\cdot10^{-8}$ & $0.059 \pm 0.002 $ & 17 \\ 
0.93 & 1.05 & $(1.06 \pm 0.11)\cdot10^{-8}$ & $0.039 \pm 0.002 $ & 11 \\ 
1.05 & 1.17 & $(6.33 \pm 1.01)\cdot10^{-9}$ & $0.028 \pm 0.002 $ & 22 \\ 
\hline
\, \\
\end{tabular}
Note: Column description. 1 and 2: Inner and outer bin radii in units of $r_{200}$; 3: Emission measure rescaled by $\Delta_{SSC}$ in units of cm$^{-6}$ Mpc; 4: Average proton density in units of $10^{-3}$ cm$^{-3}$; 5: Scatter of the various profiles relative to the mean value in percent.
\end{table}
\end{center}

\subsection{Gas mass}

We computed the gas mass from our deprojected density profiles and stacked them in the same way as described above. In the self-similar model, the gas mass is expected to follow the relation $M\propto T^{3/2}$ \citep[e.g.,][]{bryan}. However, observational works indicate that the actual $M_{gas}-T$ relation is steeper than the expected self-similar scaling \citep{neumann01,arnaud07,croston} because of the lower gas fraction in groups and poor clusters. For this work, we use the relation determined from the REXCESS sample \citep{croston} to rescale our gas mass profiles,
\begin{equation} M_{gas} \propto E(z)^{-1} \left(\frac{kT}{10\mbox{ keV}}\right)^{1.986}. \end{equation}

\begin{figure}
\resizebox{\hsize}{!}{\includegraphics{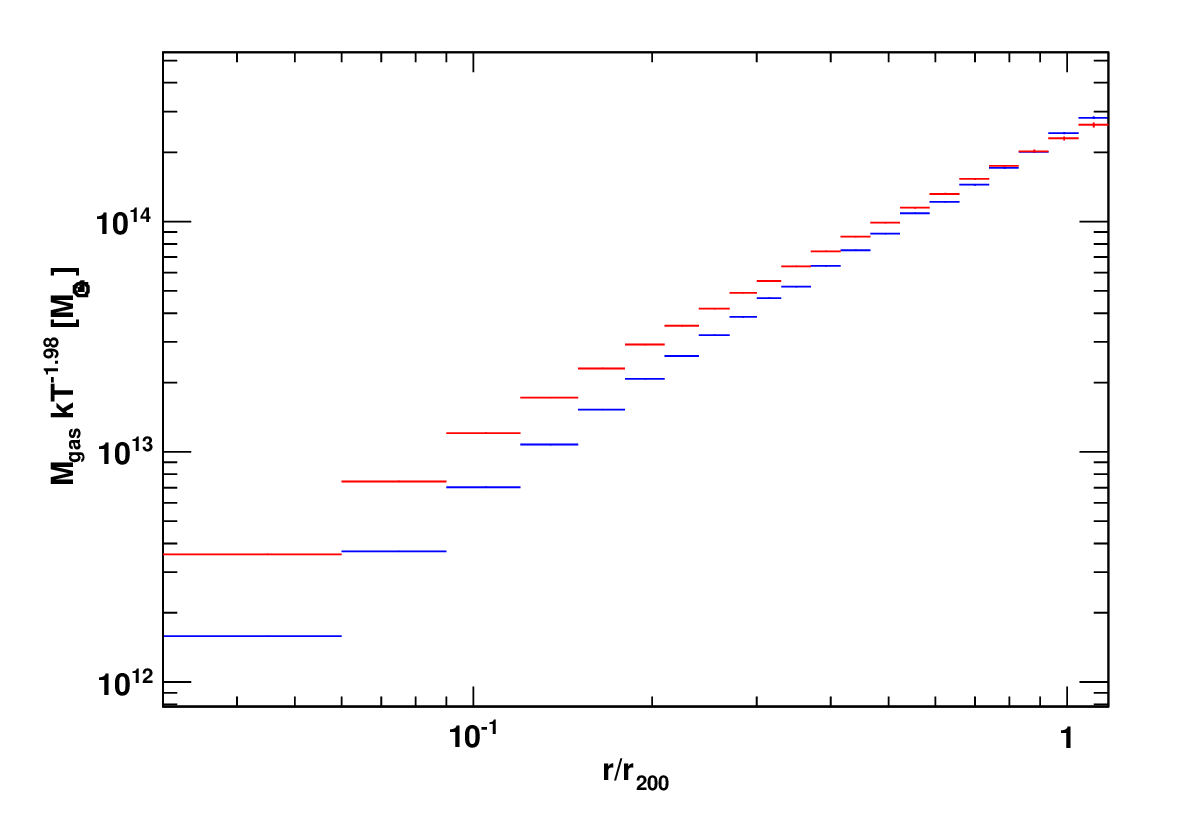}}
\caption{Enclosed gas mass profiles for CC (red) and NCC systems (blue). The data were rescaled by $E(z)kT^{-1.986}$ as observed in the REXCESS sample \citep{croston}.}
\label{mgas}
\end{figure}

As above, we divided the sample into CC and NCC classes, and stacked the two classes individually. In Fig. \ref{mgas} we show the mean gas mass profiles for CC and NCC clusters. As expected, CCs have a higher gas mass in their inner regions, since their central densities are higher. More interestingly, we see that the two profiles converge in cluster outskirts, and exhibit a gas mass around the virial radius that is consistent within the error bars. At $r_{200}$, the universal gas mass is

\begin{equation}M_{gas,200}=(2.41\pm0.05)\times10^{14} E(z)^{-1} \left(\frac{kT}{10\mbox{ keV}}\right)^{1.986} M_\odot, \end{equation}

\noindent with a scatter of 17\% around the mean value. This result follows from the higher density measured in average beyond $\sim 0.3r_{200}$ in NCC clusters and the steeper slope of CC profiles in the outskirts (see Sect. \ref{secem}). The lower density of CC clusters in the outer regions compensates for the well-known excess observed in the cores, such that the total gas mass contained within the dark-matter halo follows a universal relation. We also estimated the average gas fraction by computing the expected value of $M_{200}$ using the scaling relations of \citet{arnaud}. For our sample, we find a mean gas fraction within $r_{200}$ of

\begin{equation}f_{gas,200}=(0.15\pm0.01)\left(\frac{kT}{\mbox{10 keV}}\right)^{0.486},\end{equation}

\noindent in good agreement with previous works \citep[e.g.,][]{vikhlinin06,mccarthy07}, which for the most massive objects corresponds to $\sim89\%$ of the cosmic baryon fraction \citep{wmap7}.

\subsection{Azimuthal scatter}
\label{secscat}

Following the method described in Sect. \ref{azmet}, we computed the azimuthal scatter of the surface-brightness profiles for all the clusters in our sample, and rescaled the scatter profiles by our estimated value of $r_{200}$. We then stacked the profiles using the same procedure as described above and computed the mean azimuthal scatter. We recall that since the surface brightness depends on $n_e^2$, the variations in density are less important than the ones computed here.

\begin{figure}
\resizebox{\hsize}{!}{\includegraphics{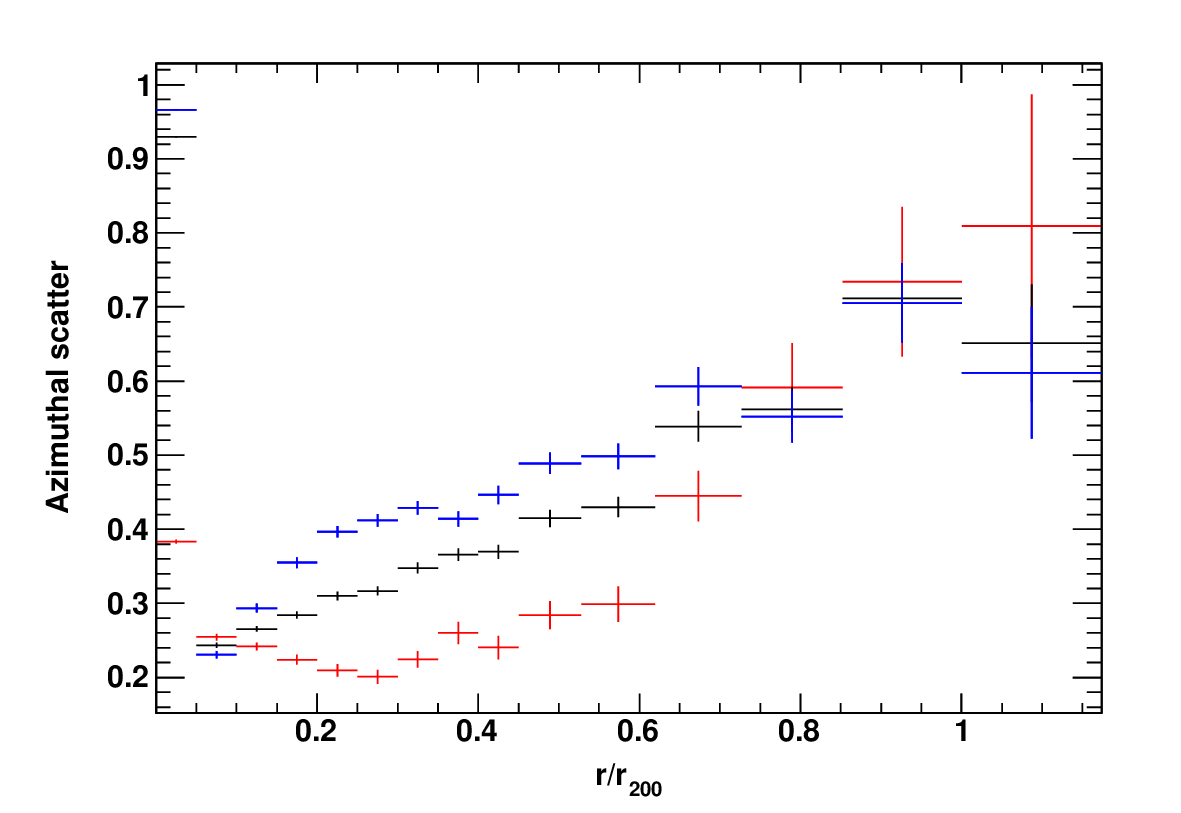}}
\caption{Stacked azimuthal scatter in surface-brightness for the entire cluster sample (black). The red and blue data represent the mean profile extracted from CC and NCC clusters, respectively.}
\label{mscat}
\end{figure}

In Fig. \ref{mscat} we plot the average scatter profile, compared to the mean value for CC and NCC clusters. The increase in the innermost bin is an artifact introduced by the small number of pixels in the center of the images, so it should be neglected. At small radii ($r<0.5r_{200}$) we find a clear difference between CC and NCC clusters, which is easily explained by the more disturbed morphology of the latter. In this radial range, CC profiles exhibit a scatter of 20-30\%, which corresponds to density variations on the order of 10\%, in good agreement with the value predicted by \citet{vazzascat} from numerical simulations. Conversely, beyond $r\sim r_{500}$, the profiles for CC and NCC clusters are similar, and indicate a high scatter value (60-80\%). 

We investigated whether any systematic effect could affect our result in cluster outskirts, where the background is dominating with respect to the source. Indeed, in such conditions, the total scatter is dominated by the statistical scatter. In case the mean level of systematic uncertainties in the CXB reconstruction exceeds our adopted value of 6$\%$, Eq. \ref{statscat} immediately implies that the intrinsic scatter would be overestimated. The presence of both intrinsic and statistical scatter could also introduce some covariance term, which is not taken into account in Eq. \ref{statscat}. To test this hypothesis, we ran a set of simulations including source and background, where we introduced a given level of intrinsic scatter for the source and a systematic error in addition to the Poisson statistics for the background. We then computed the intrinsic level of scatter following Eq. \ref{statscat}. Our simulations indicate that, even when increasing the level of systematic uncertainties to 12\% of the CXB value, a significant bias in the measured scatter only appears when the source-to-background ratio is close to the systematic uncertainties. Since, by construction, we never detect any signal when the source is less than $\sim$15\% of the CXB value, our results are unaffected by these effects, and we can conclude with good confidence that the high level of scatter measured beyond $\sim r_{500}$ is an intrinsic property of our cluster sample.

In addition, we also tested whether the scatter for the two populations in the outermost regions could be affected by small-number statistics or driven by some particular objects. Of the 31 objects in our sample, a measurement of the scatter at $r_{200}$ could be obtained for 23 of them (12 NCC and 11 CC). We used a jackknife method to test whether a single object dominates the results for any of the two populations; i.e., we randomly exclude one or two profiles from the sample, recompute the mean profiles, and examine the distribution of the mean values. In both cases, the distribution of results is regular, which indicates that our results are not biased by a particular object.

V99 also investigated the deviations from spherical symmetry by measuring the value of $\beta$ in six sectors in the radial range $r>0.3r_{180}$, and concluded that the assumption of spherical symmetry is relatively well satisfied in cluster outskirts, at variance with our results (see Fig. \ref{mscat}). However, when fitting a $\beta$-model the fit is mostly driven by the shape of the profile in the innermost region, where the statistics are higher. Conversely, our method is model-independent, and it directly stacks the data at similar radii. For relaxed objects, our data also indicate little deviation from spherical symmetry at $r<r_{500}$, and a significant scatter is only observed beyond $r_{500}$, so it is probable that these deviations would not be reflected in the $\beta$-model fit. For instance, the case of A2029 is striking. While, in agreement with V99, we find little azimuthal variations in $\beta_{outer}$, we observe a high level of scatter in this object beyond $r_{500}$, which is explained by the presence of a possible filament connecting A2029 to its neighbor A2033 in the north (see \citet{gasta10} and Appendix \ref{appendix}). Moreover, V99 deliberately excluded a number of systems with obviously disturbed morphologies, such as A3558 and A3266, which we included in our sample. Therefore, our results do not contradict those of V99.

\section{Comparison with numerical simulations}
\label{secsim}

In this section, we compare our observational results with three different sets of numerical simulations \citep{roncarelli,nagai07,vazza10}. We analyze the results of a composite set of cosmological runs, obtained by the different authors with slightly different cosmological and numerical setups. In addition, the preliminary data reduction was made on each dataset following independent post-processing techniques, aimed at assessing the role of gas clumping on the comparison between simulated mock and real X-ray observations. Our aim in this project is to test the most general and converging findings of such different runs against our observations with \emph{ROSAT}/PSPC.

\subsection{Simulations}

\subsubsection{{\tt ENZO}}

We use a sample of 20 simulated clusters from the high-resolution and non-radiative (NR) resimulations of massive systems presented in \citet{vazza10}. In this set of simulations, adaptive mesh refinement in the {\tt ENZO} 1.5 code \citep{norman07} has been tailored to achieve high resolution in the innermost regions of clusters (following the increase in gas and DM overdensity), and also in the outermost cluster regions, following the sharp fluctuations of the velocity field, associated with shocks and turbulent motions in the ICM. For a detailed presentation of the statistical properties of the thermal gas (and of turbulent motions) in these simulated systems we refer the reader to \citet{vazza10,vazza11a}. \\

\subsubsection{{\tt ART}}

We analyze a sample of ten simulated clusters with $T_X>2.5$ keV from the sample presented in \citet{nagai07b,nagai07}. These simulations are performed using the adaptive refinement tree ({\tt ART}) N-body+gas-dynamics code \citep{kravtsov99,kravtsov02}, which is a Eulerian code that uses adaptive refinement to achieve high spatial resolution (a few kpc) in self-consistent cosmological simulations. To assess the impact of cluster physics on the ICM properties, we compared two sets of clusters simulated with the same initial conditions but with different prescription of gas physics. In the first set, we performed hydrodynamical cluster simulations without gas cooling and star formation. We refer to this set of clusters as NR clusters. In the second set, we turned on the physics of galaxy formation, such as metallicity-dependent radiative cooling, star formation, supernova feedback, and a uniform UV background. We refer to this set of clusters as cooling+star formation (CSF) clusters. For detailed descriptions of the gas physics and mock X-ray images we refer the reader to \citet{nagai07b,nagai07}. 

Following \citet{nagai}, we also computed the clumping-corrected gas density profiles of X-ray emitting gas with $T>10^6$~K for comparisons with X-ray observations. Indeed, the formation of dense clumps increases the emissivity of the gas, which leads to an overestimation of the measured gas density when the assumption of constant density in each shell is made. For these profiles, we computed the average squared density from the simulations in each radial bin and took the square root of the total to mimic the reconstruction of density profiles from real data \citep[see][for details]{nagai}. 

\subsubsection{{\tt GADGET}}

This set includes four massive halos simulated with the {\tt GADGET-2} Tree-SPH code \citep{springel}, with $M_{200} > 10^{15} M_\odot$ \citep[for a detailed description see][and references therein]{roncarelli}. Each object was simulated following two different physical prescriptions: a NR run \citep[referred to as \textit{ovisc} in ][]{roncarelli} and a run including cooling, star formation, and supernovae feedback (CSF).

To eliminate the dense clumps that dominate the density and surface brightness in the outskirts, when computing the profiles for every radial bin, we excise the one per cent of the volume that corresponds to the densest SPH particles. This empirical method mimics the procedure of masking bright isolated regions from the analysis of observed clusters.

\subsection{Comparison of gas density profiles}

\begin{figure*}
\resizebox{\hsize}{!}{\hbox{\includegraphics{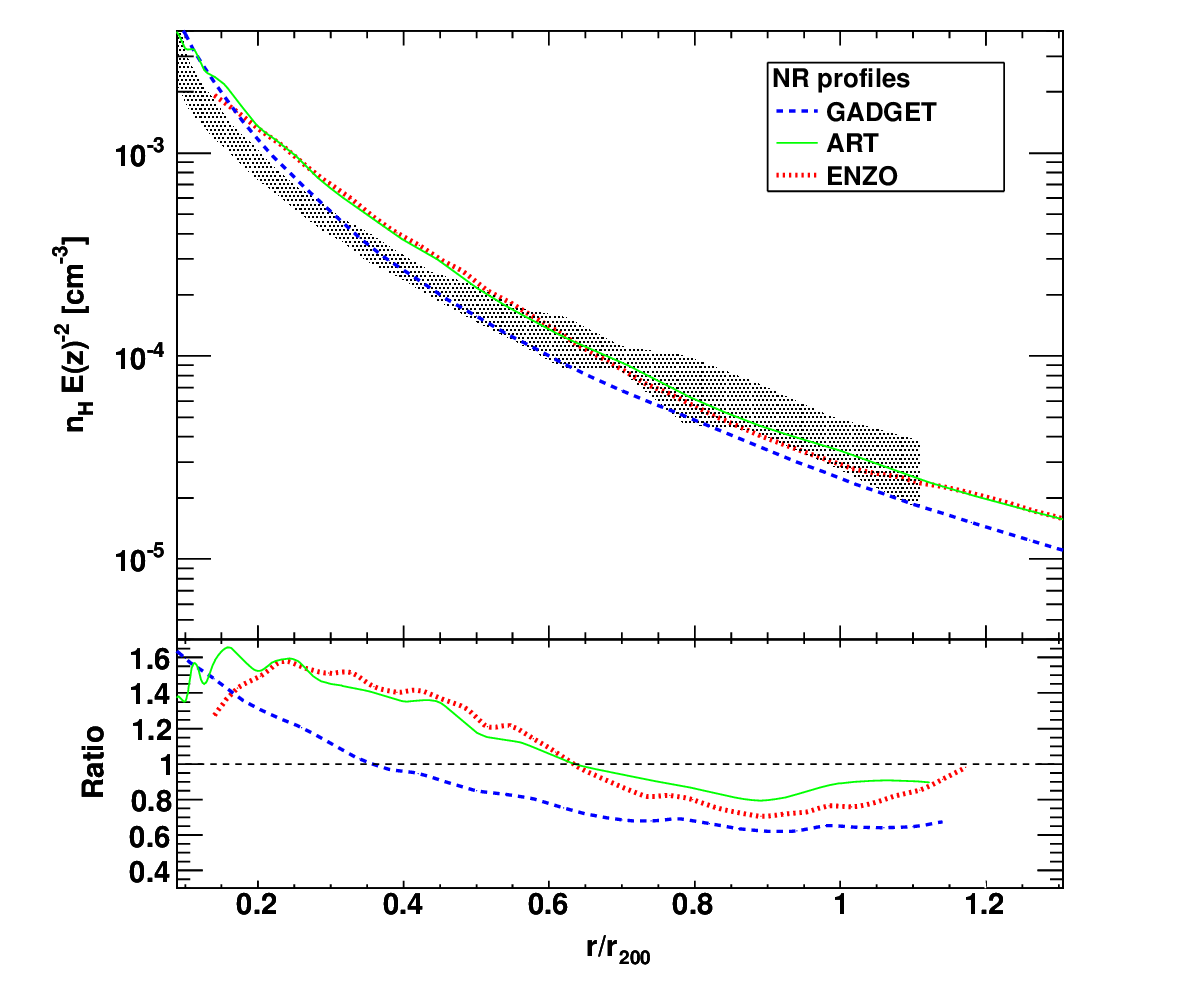}\includegraphics{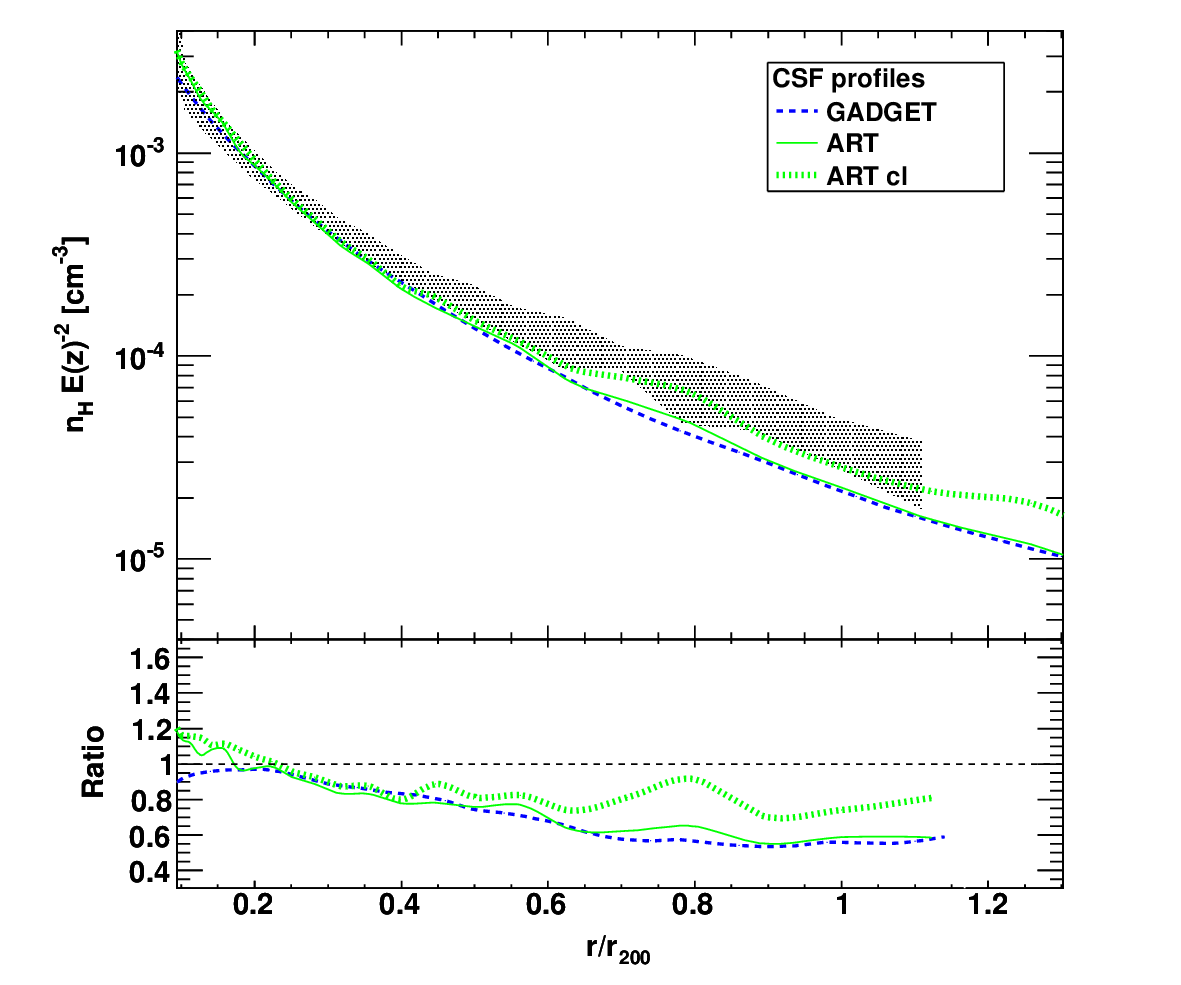}}}
\caption{Comparison between the mean \emph{ROSAT} density profile for our sample and the different sets of numerical simulations. The shaded area indicates the data and $1\sigma$ scatter as shown in Fig.  \ref{meandens}. The bottom panels show the ratio between simulations and data as a function of radius. \textit{Left:} Comparison with NR simulations. The dotted red curve represents the {\tt ENZO} profile \citep{vazza10}, the solid green curve shows the {\tt ART} simulations \citep{nagai07}, and the dashed blue curve is the {\tt GADGET} profile \citep{roncarelli}. \textit{Right:} Same with CSF simulations. The dashed blue line shows the {\tt GADGET} simulations, while the green curves show the {\tt ART} profiles, for the total density (solid) and corrected for clumping \citep[dotted,][]{nagai}.}
\label{denssimnr}
\end{figure*}

We compared the simulations with our observed mean \emph{ROSAT} density profile (see Fig. \ref{meandens} and Table \ref{tabmean}). We present the detailed comparison in Fig. \ref{denssimnr}, with the NR simulations (left hand panel) and with the CSF simulations (right). From the figures, we find relatively good agreement between all the different sets of simulations, especially beyond $\sim0.7r_{200}$. The NR {\tt GADGET} run has a lower normalization than the corresponding grid codes, because in {\tt GADGET} the fraction of baryons virializing into clusters is less than the cosmic value ($\sim78\%$ of the cosmic baryon fraction), while grid codes predict a baryon fraction in clusters very close to the cosmic value. In general, we see that the predicted density profiles are too steep compared to the data. We note that NR runs predict steeper profiles than the runs including cooling, star formation, and feedback effects. CSF profiles also have lower normalizations, since radiative cooling transforms a fraction of the gas into stars. The profile including the effects of clumping shows the best agreement with the data.

To quantify this effect, we fitted the various profiles in three different radial ranges ($0.2-0.4r_{200}$, $0.4-0.65r_{200}$, and $0.65-1.2r_{200}$). In the inner regions, the effects of additional physics are expected to be important, thus highlighting the differences between NR and CSF runs. The radial range $0.4-0.65r_{200} (\approx 0.6-1r_{500})$ is a good range for comparing with the data, since the effects of radiative cooling should be small, and data from several different satellites are available for cross-check. On the observational side, the density profiles in this radial range are well-fitted by the $\beta$-model (see Eq. \ref{beta}), and several independent works converge to the canonical value of $\beta\sim0.7$ \citep[e.g.,][]{mohr,ettori99,vikhlinin99,croston,ettbal,eckert}. As a benchmark, we computed the values of $\beta$ for our average density profile and the various sets of simulations, fixing the core radius to $0.12r_{200}$ \citep[e.g.,][]{mohr}. The results of this analysis are shown in Table \ref{betavalues}. The fits to the observational data were performed on the emission-measure profiles (see Sect. \ref{secem}) to take advantage of the larger number of bins and minimize the uncertainties linked to the deprojection procedure. 


\begin{table}
\begin{center}
\caption{ \label{betavalues} Values of the $\beta$ parameter \citep{cavaliere} in several radial ranges for the average \emph{ROSAT} profiles and the various sets of simulations.}
\begin{tabular}{lccc}
\hline
Data set & $\beta_{0.2-0.4}$ & $\beta_{0.4-0.65}$ & $\beta_{0.65-1.2}$ \\
\hline
\hline
Data, total & $0.661 \pm 0.002$ & $0.710\pm0.009$ & $0.890\pm0.026$ \\
Data, CC & $0.700\pm0.004$ & $0.699\pm0.016$ & $1.002\pm0.057$ \\
Data, NCC & $0.635 \pm 0.003$ & $0.723\pm0.011$ & $0.853\pm0.029$ \\
{\tt ENZO} & 0.744 & 0.945 & 0.952 \\
{\tt ART}, NR & 0.801 & 0.956 & 0.983 \\
{\tt ART}, CSF & 0.808 & 0.842 & 1.005 \\
{\tt ART}, NR, cl & 0.701 & 0.824 & 0.854 \\
{\tt ART}, CSF, cl & 0.803 & 0.718 & 0.902 \\
{\tt GADGET}, NR & 0.856 & 0.857 & 0.971 \\
{\tt GADGET}, CSF & 0.756 & 0.864 & 0.944 \\
\hline
\, \\
\end{tabular}
Note: The core radius was fixed to $0.12r_{200}$ in all cases. The subscript \textit{cl} indicates the profiles corrected for the effect of clumping using the method described in \citet{nagai}.
\end{center}
\end{table}

These numbers confirm the visual impression that the simulated gas density profiles are steeper than the observed ones. In the $0.4-0.65r_{200}$ range, while all our datasets converge to a $\beta$ value very close to the canonical value, all the simulations lead to significantly steeper gas profiles, with $\beta$ values higher than 0.85, with the exception of the {\tt ART} profile that includes CSF and clumping. Therefore, we can see that at this level of precision the effects of additional physics cannot be neglected, even in regions well outside of the cluster core.

\begin{figure*}
\resizebox{\hsize}{!}{\hbox{\includegraphics{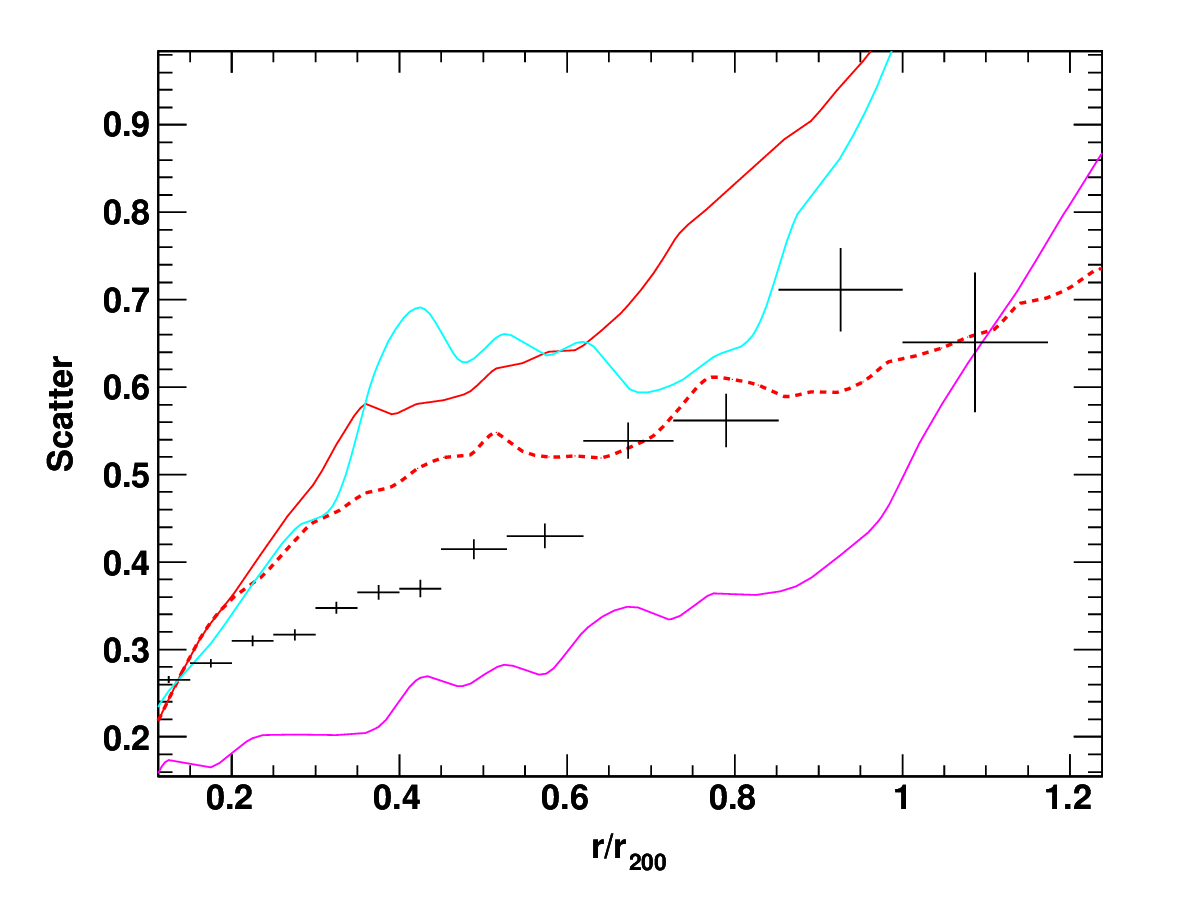}\includegraphics{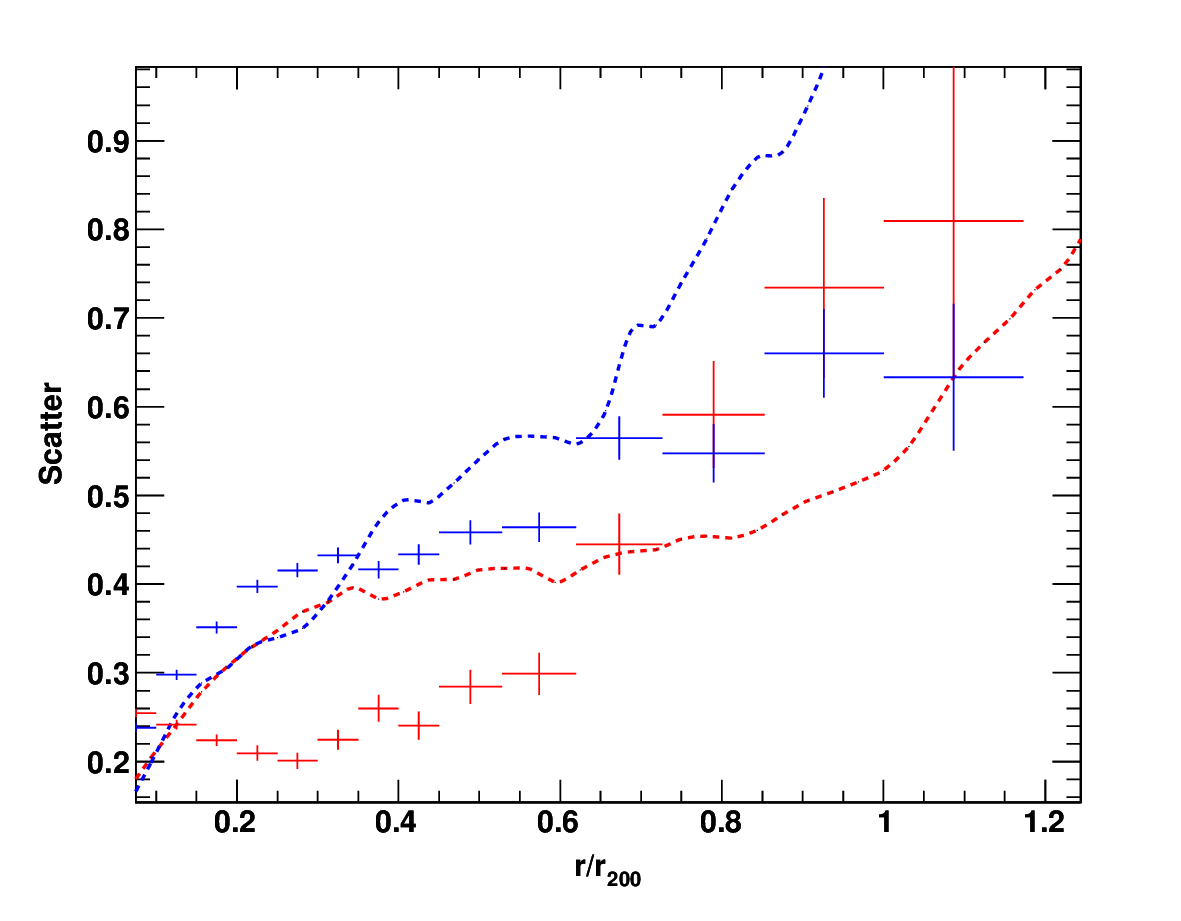}}}
\caption{\textit{Left:} Comparison between the average observed azimuthal scatter profile from Fig. \ref{mscat} (black) and the scatter in the simulations for the {\tt ENZO} runs (red), for the total scatter (solid line) and when filtering out the 1\% most-luminous cells (dashed curve). The cyan (NR) and magenta (CSF) curves represent the scatter in the {\tt ART} simulations. \textit{Right:} Same for the CC (red) and NCC (blue) observed profiles, compared to the 1\%-filtered {\tt ENZO} profiles for the morphologically relaxed (red) and disturbed (blue) simulated clusters. }
\label{simscat}
\end{figure*}

The results presented in Table \ref{betavalues} also highlight the differences between NR and CSF runs. Inside $r_{500}$, the simulations including additional physics lead to flatter density profiles compared to the NR runs. In this case, gas cooling converts a fraction of the X-ray emitting gas into stars. Since the cooling efficiency decreases with radius, more gas disappears from the X-ray range in the central regions, which results in flatter density profiles and lower normalizations. We note, however, that this effect is probably overestimated in the CSF simulations. Indeed, it is well-known that these simulations predict a stellar fraction that is well above the observed value \citep[e.g.,][]{kravtsov05,borgani}. This effect is particularly strong in the {\tt ART} CSF simulation, for which nearly one third of the gas is converted into stars. Beyond $r_{500}$, there is little difference between NR and CSF runs; i.e., the effects of additional physics are not important. At large radii, the effect of gas clumping \citep{nagai} dominates and flattens the observed profiles. As we can see in Table \ref{betavalues} and in the right hand panel of Fig. \ref{denssimnr}, the {\tt ART} profile including both additional physics and a post-processing treatment of clumping reproduces the behavior of the data more closely, even though it is still slightly too steep.

\subsection{Azimuthal scatter}

A study of the azimuthal scatter in the radial profiles of density, temperature, entropy and X-ray brightness of simulated {\tt ENZO} clusters has been presented in \citet{vazzascat}. In this case, we differ from the analysis reported there by computing the azimuthal scatter from more angular sectors, N=12, than for N=2, 4, and 8 explored in \citet{vazzascat}. In the simulations, several dense clumps are present, which may bias the predicted scatter. To overcome this problem, we computed the scatter of the simulated clusters both for the total gas distribution and by filtering out the 1\% most X-ray luminous cells, as in \citet{roncarelli}, which removes a large fraction of the clumps. 

We also performed a similar analysis on the set of {\tt ART} simulations, both for the NR and CSF runs. In this case, we analyzed mock X-ray images using the same method as the observational data (see Sect. \ref{azmet}), and applied our point-source detection algorithm to remove the most prominent clumps.\footnote{Because of how few objects are considered, we ignored the {\tt GADGET} simulations for this analysis. For a comparison between {\tt GADGET} and {\tt ENZO} scatter profiles, we refer the reader to \citet{vazzascat}.}

In Fig. \ref{simscat}, we show the measured scatter profile from Fig. \ref{mscat}, together with the scatter profiles of X-ray brightness from {\tt ENZO} and {\tt ART} simulations. Interestingly, we note that NR runs (red and cyan) overestimate the observed azimuthal scatter, while CSF simulations underestimate it. In the latter case, radiative cooling lowers the entropy of the gas, which makes it sink into the cluster's potential well. This effect produces more spherical X-ray morphologies, thus lowering the azimuthal scatter. Conversely, in NR runs, the effects of dynamics are more important, which create more substructures and increases the azimuthal scatter.

Interestingly, the profile that best reproduces the data is the {\tt ENZO} profile for which the 1\% most-luminous pixels were filtered out. This may indicate that some clumps are indeed present in the observations, but were detected as point sources and were masked for the analysis. We note that, even if in this case the azimuthal scatter from NR simulation runs is in good agreement with the \emph{ROSAT} data, the absolute profiles of density are too steep compared to observations (see the left hand panel of Fig. \ref{denssimnr}). However, our definition of the azimuthal scatter (Eq. \ref{stot}) is normalized to the absolute value of the profile at each radii, which makes it a rather robust proxy of cluster asymmetries on large $\sim$ Mpc scales. 

In the right hand panel of Fig. \ref{simscat}, we also show the average radial trends of the azimuthal scatter for the projected X-ray emission from the {\tt ENZO} clusters after dividing the dataset into 11 CC-like and 9 NCC-like objects, compared to the observed scatter profiles for the CC and NCC classes from Fig. \ref{mscat}. This division is of course only qualitative, since no radiative cooling is modeled in these runs. However, our sample can be divided into classes that are quite similar to observed CC and NCC properties, based on the analysis of the power ratios $P_3/P_0$ and of the centroid shift {\it w}, evaluated within $r_{500}$ as in \citet{cassano}. We classify as NCC-like systems those for which the values of $P_3/P_0>10^{-7}$ and $w>0.02$ were found in at least two of the three projected maps along the coordinate axes, or as CC-like otherwise, identical to what was done for the same sample in \citet{vazza11a}. 

In this figure, we can clearly see that the radial trend of the difference between the two populations disagrees. While in simulations the two trends detach as we move farther out in the cluster atmospheres, in the observed profiles the most prominent differences are found in the range $0.2 \leq r/r_{200}\leq 0.8$. In the CC case, we find better qualitative agreement in the outskirts than in the central regions. This is not surprising, given that radiative cooling and energy feedback from central AGNs are missing in these runs. Indeed, as we can see in the left hand panel of Fig. \ref{simscat}, radiative cooling has a strong impact on the general morphology of clusters \citep{fang,lau11}.  On the other hand, the simulated disturbed systems have a larger scatter in the outskirts than the observed NCC clusters. However, we observe large differences in the scatter between the various NCC profiles, such that the result may be affected by small-number statistics. In any case, since the selection criteria are very different, we do not expect a one-to-one correlation between the various classes. 

\section{Discussion}
\label{secdisc}

\subsection{Observational results}

In agreement with earlier works using \emph{ROSAT} (V99, N05) and \emph{Chandra} \citep{ettbal}, but at variance with some recent results from \emph{Suzaku} \citep{bautz,simionescu,george} and \emph{XMM-Newton} \citep{norbi}, our analysis reveals that on average the slope of the density profiles steepens beyond $r_{500}$ (see Table \ref{betavalues}). This result indicates that the latter results may have been performed along preferential directions connected with the large-scale structure (e.g., in the direction of filaments). Indeed, the narrow FOV of \emph{Suzaku} only allowed sparse coverage of the outskirts of nearby clusters, so that these measurements might be the result of azimuthal variations. In the case of A1795, \citet{bautz} detected a significant signal only in the northern direction, while the Perseus result \citep{simionescu} was obtained along two narrow arms, covering less than 10\% of the cluster's extent at $r_{200}$. Moreover, using several offset \emph{ROSAT}/PSPC pointings of the Perseus cluster, \citet{ettoriper} observed clear azimuthal variations in the density and gas fraction. Therefore, it is likely that the aforementioned measurements are not representative of the cluster as a whole. This picture is supported by our analysis of azimuthal variations in cluster outskirts, which suggests that even CC clusters exhibit significant departures from spherical symmetry around $r_{200}$. Consequently, a full azimuthal coverage is required to study the global behavior of cluster outer regions.

An important result of this work is the systematic difference between CC and NCC cluster populations observed beyond $\sim 0.3r_{200}$ (see Fig. \ref{meanem}). As explained in Sect. \ref{meanem}, this effect seems to be an intrinsic difference between the two classes, since it is does not correspond to a biased distribution of our sample in temperature or redshift. Our scaled gas mass profiles provide a natural explanation for this result (see Fig. \ref{mgas}). Indeed, when the appropriate scaling is applied, the steeper density profiles of CCs in the outskirts compensate exactly for the excess density in the central regions, such that clusters with the same virial mass have the same gas mass enclosed within $r_{200}$, albeit distributed in a different way for relaxed and disturbed objects. This result was expected in the old cooling-flow scenario \citep{fabian}, in which radiative cooling causes the gas to flow inwards and accumulate in the central regions. While in the central regions AGN feedback prevents the gas from cooling below a certain level \citep[e.g.,][]{mcnamara}, the entropy injected by the central AGN is not strong enough to balance the flow in the outer regions of clusters, which explains the steep density profiles seen in Fig. \ref{meanem}. Conversely, merging events are capable of injecting a very large amount of energy in the ICM, which results in an efficient redistribution of the gas between the core and the outer regions and creates the flatter density profiles measured for NCC clusters.

We also determined the typical scatter in surface-brightness as a function of radius (see Fig. \ref{mscat}) and split the data into the CC and NCC classes. In the central regions, we observe a systematic difference between CC and NCC clusters, with NCC clusters showing a higher level of scatter than CC. This result is easily explained by the larger number of substructures generally observed in NCC clusters \citep[e.g.,][]{sanderson}. For CC clusters, we measure a scatter of $20\%-30\%$ below $0.5r_{200}$, which corresponds to small variations ($\sim10\%$) in gas density. This indicates that the azimuthal scatter in the inner regions ($r<0.5r_{200}$) can be used to estimate the X-ray state of clusters, as suggested by \citet{vazzascat}. Conversely, the scatter of CC profiles increases in cluster outskirts, and there is no observed difference between the two classes. Interestingly, we note that for CC clusters the turnover in Fig. \ref{mscat} occurs around $r_{500}$, which coincides with the radius beyond which large scale infall motions and filamentary accretions are generally non-negligible \citep[e.g.,][]{evrard96}. Inside $r_{500}$, the gas is virialized in the cluster's potential well, and it shows only little deviations from spherical symmetry. Beyond $r_{500}$, accretion processes are important, and the gas is located mostly along preferential directions (i.e., filaments). As a result, the distribution of the gas becomes strongly anisotropic, even for clusters that exhibit a relaxed morphology in their inner regions.

\subsection{Comparison with simulations}

Comparing our density profiles with numerical simulations, we find that all NR simulations predict very steep profiles already starting from $\sim0.2r_{200}$, with values of the $\beta$ parameter greater than 0.85 in the $0.4-0.65r_{200}$ range (see the left hand panel of Fig. \ref{denssimnr} and Table \ref{betavalues}). This indicates that including non-gravitational effects is needed to reproduce the observed slope, even well outside of cluster cores. The runs including additional physics are in better qualitative agreement with the observations (see the right hand panel of Fig. \ref{denssimnr}), although their gas fraction is too low because of overcooling ($\sim10\%$ compared to $\sim15\%$). However, it seems unlikely that star formation and galactic winds (as in the CSF runs explored here) are the only feedback mechanisms needed to reproduce observed clusters. Indeed, simple feedback models still face severe problems in matching the properties of the stellar components inside galaxy clusters, as well as the properties of galaxies within them \citep[e.g.,][for a recent review]{borgani}.

As illustrated in Table \ref{betavalues}, gas clumping may also play a role in reconciling simulations with observations. Indeed, if an important fraction of the gas in cluster outskirts is in the form of dense gas clumps, as suggested in simulations \citep{nagai}, the emissivity of the gas would be significantly increased, thus leading to an overestimation of the gas density when the assumption of constant density in each shell is made. Our results show that the treatment of gas clumping slightly improves the agreement between data and simulations (see the right hand panel of Fig. \ref{denssimnr}). In addition, gas clumping also provides an alternative interpretation for our observed difference between the CC and NCC populations beyond $0.3r_{200}$. Indeed, simulations predict a larger clumping factor in unrelaxed clusters compared to relaxed systems for the same average density, which would result in a higher observed density in the former. At the moment, it is not clear whether this difference is caused by gas redistribution or clumping, or if both of these effects play a role to some extent.

On the other hand, we find that numerical simulations can reproduce qualitatively the observed azimuthal scatter in the galaxy cluster gas density profiles (see Fig. \ref{simscat}), although they fail to reproduce the trends observed for the CC and NCC populations separately. Interestingly, we find that the observed azimuthal scatter is reproduced with reasonable accuracy when the 1\% most luminous clumps are filtered out, whereas the NR simulations with no filtering overestimate the observed level of azimuthal scatter at all radii. Two possible interpretations can be put forward to interpret this result. Observationally, it is possible that the dense clumps were detected as point sources and were filtered out of our observations. If this is the case, long exposures with high-resolution X-ray telescopes (\emph{Chandra} or \emph{XMM-Newton}) should allow us to characterize the point sources and distinguish between dense clumps and background AGN, possibly unveiling the population of accreting clumps in cluster outskirts. Conversely, if such observations do not confirm the existence of the clumps, it would imply that NR simulations significantly overestimate the amount of clumping in cluster outskirts, which would weaken the case for the interpretation recently put forward to explain the flattening of the entropy profiles observed in a few cases \citep{simionescu,norbi}. 

As shown in Fig. \ref{simscat}, radiative cooling may also help reconcile the NR simulations with the data. Indeed, radiative cooling lowers the entropy of the gas and makes it sink into the potential well, which produces clusters with more spherical morphologies \citep{lau11} and thus reduces the azimuthal scatter. Since we know that this effect is overestimated in our CSF simulations, radiative cooling likely reduces the azimuthal scatter with respect to NR simulations, although not as much as what is predicted here. This effect may also explain why NR simulations fail to reproduce the average scatter profiles of CC clusters (see the right hand panel of Fig. \ref{simscat}).

Alternatively, AGN feedback may be an important ingredient that is rarely taken into account in numerical simulations. Recently, \citet{pratt10} observed an anti-correlation between entropy and gas fraction, such that multiplying cluster entropy profiles by the local gas fraction allows recovery of the entropy profiles predicted from adiabatic compression; i.e., the excess entropy observed in cluster cores is balanced by a lower gas fraction, and the total entropy follows the predictions of gravitational collapse. \citet{mathews} interpret this result in terms of the total feedback energy injected in the ICM through various giant AGN outbursts, which they estimate to be as large as $10^{63}$ ergs. In this scenario, feedback mechanisms are preventing the gas from collapsing into the potential well, causing a deficit of baryons in the inner regions of clusters, hence flattening the observed density profiles. Moreover, it is well known that this mechanism also takes place on group and galaxy scales, leading to shallower density profiles in the accreting clumps. As a result, the gas distribution in cluster outskirts would be more homogeneous than predicted in NR simulations, in agreement with our observed azimuthal scatter profiles. Therefore, although its implementation into numerical simulations is challenging \citep{sijacki}, AGN feedback could be an important effect for reconciling simulations with observations. A more complex picture of the ICM, possibly including the detailed  treatment of magnetic fields, cosmic rays, and thermal conductions (and of the instabilities arising from these ingredients), would still represent a challenge for current cosmological simulations. 

\section{Conclusion}

In this paper, we have presented our analysis of a sample of local ($z=0.04-0.2$) clusters with \emph{ROSAT}/PSPC, focusing on the properties of the gas in cluster outskirts. We then compared our observational results with numerical simulations \citep{roncarelli,nagai,vazzascat}. Our main results can be summarized as follows.

\begin{itemize}
\item
We observed a general trend of steepening in the radial profiles of emission-measure and gas density beyond $\sim r_{500}$, in good agreement with earlier works from \citet{vikhlinin99}, \citet{neumann05}, and \citet{ettbal}. As a result, the shallow density profiles observed in several clusters by \emph{Suzaku} \citep{bautz,simionescu} are probably induced by observations in preferential directions (e.g., filaments) and do not reflect the typical behavior of cluster outer regions.

\item
We found that NCC clusters have in average a higher density than CC systems beyond $\sim0.3 r_{200}$, which cannot be easily explained by any selection effect. We interpreted this result by a different distribution of the gas in the two populations: the well-known density excess in the core of CC clusters is balanced by a slightly steeper profile in the outskirts, which leads to the same gas mass enclosed within $r_{200}$ in the two populations (see Fig. \ref{mgas}). Alternatively, this result could be caused by a larger clumping factor in disturbed objects, leading to an overestimate of the gas density of NCC clusters in the external regions.

\item
We also observed that NCC systems have higher azimuthal scatter than CCs in the central regions, which is easily explained by the more disturbed morphology of NCC clusters. Conversely, beyond $\sim r_{500}$ both populations show a similar level of asymmetry (60-80\%), which suggests that a significant fraction of the gas is in the form of accreting material from the large-scale structures. 

\item
Comparing our \emph{ROSAT} density profile with numerical simulations, we found that all NR numerical simulations fail to reproduce the observed shape of the density profile, predicting density profiles that are significantly too steep compared to the data (see Table \ref{betavalues} and Fig. \ref{denssimnr}). This implies that nongravitational effects are important well outside the core region. The runs including additional physics (cooling, star formation, SN feedback) predict flatter profiles, although still too steep compared to the observations. Besides, it is well known that these simulations overpredict the stellar fraction in clusters \citep{borgani}. A slightly better agreement is found when a treatment of the observational effects of gas clumping is adopted \citep{nagai}.

\item
NR simulations are able to predict the observed azimuthal scatter profile with reasonable accuracy, but only when the 1\% most luminous cells are filtered out (see Fig. \ref{simscat}). This result implies that either (i) the clumps are quite bright and were masked as point sources in our analysis pipeline, in which case offset \emph{XMM-Newton} and \emph{Chandra} observations will be able to characterize them spatially and spectrally, or (ii) the non-radiative simulations significantly overestimate the effects of clumping on the observable X-ray properties. Because of the absence of cooling, it is however hard for these simulations to reproduce the observed trends of azimuthal scatter for the two populations (CC and NCC) separately.

\end{itemize}

As an alternative explanation, we suggest that AGN feedback might be important even at  large radii, and could help to reconcile observations and simulations. Indeed, recent works \citep{pratt10,mathews} indicate that feedback mechanisms may be responsible for the well-known deficit of baryons in cluster cores, thus leading to flatter gas distributions out to large radii. Moreover, the existence of such mechanisms on group and galaxy scales could also dilute the accreting material at large radii, leading to a smaller azimuthal scatter.

\acknowledgements{We thank Klaus Dolag for kindly prodiving the data of his {\tt GADGET} runs. SE, SM, and DN acknowledge the support from the National Science Foundation under Grant No. NSF PHY05-51164 for attending the workshop on ``Galaxy Clusters: The Crossroads of Astrophysics and Cosmology'', where part of this project has been discussed. DE was supported by the Occhialini fellowship of IASF Milano. MR and FG acknowledge the financial contribution from contracts ASI-INAF I/023/05/0, I/009/10/0 and I/088/06/0. FV acknowledges the collaboration of G.Brunetti, C. Gheller, and R.Brunino in the production of {\tt ENZO} runs studied in this work. DN was supported in part by the NSF AST-1009811, by NASA NNX11AE07G, and by the facilities and staff of the Yale University Faculty of Arts and Sciences High Performance Computing Center.}

\normalsize

\bibliographystyle{aa}
\bibliography{outskirts}

\begin{center}
\begin{landscape}
\begin{table}[hbt]
\caption{\label{master}Master table of the cluster sample. Column description: 1. Cluster name; 2. Effective exposure of the PSPC observation; 3. Redshift (from NED); 4. Hydrogen column density, $N_H$, along the line of sight \citep{kalberla}; 5. Mean temperature in the 200-500 kpc radial range; 6. $r_{200}$ from \citet{arnaud} scaling relations, in physical units; 7. Same as 6, in apparent units; 8. Central density $n_0$ (this work); 9. Central entropy $K_0$, from \citet{cavagnolo}; 10. Reference for the temperature profile (1=\citet{xmmcat}; 2=\citet{cavagnolo}; 3=\citet{sabrina}).}
\begin{tabular}{cccccccccc}
\hline
\, \\
Cluster & Exposure [ks] & $z$ & $N_H$ [$10^{22}$ cm$^{-2}$] & $kT_{200-500}$ [keV] & $r_{200}$ [kpc] & $r_{200}$ [arcmin] & $n_0$ [$10^{-3}$ cm$^{-3}$] & $K_0$ [keV cm$^2$] & Reference\\
\, \\
\hline\hline
\, \\
A85 & 10.065 & 0.05506 & 0.028 & $6.3\pm0.1$ & 1873 & 29.17 & $18.9\pm0.25$ & 12.5 & 1\\
A119 & 14.758 & 0.0442 & 0.037 & $5.0\pm0.1$ & 1673 & 32.04 & $2.1\pm0.34$ & 233.9 & 2\\
A133 & 19.429 & 0.0566 & 0.0164 & $4.0\pm0.09$ & 1494 & 22.68 & $14.0\pm0.18$ & 17.3 & 1\\
A401 & 7.519 & 0.07366 & 0.0995 & $7.9\pm0.15$ & 2077 & 24.72 & $5.3\pm0.66$ & 166.9 & 2\\
A478 & 23.019 & 0.0881 & 0.131 & $6.56\pm0.08$ & 1883 & 19.05  & $18.8\pm0.19$ & 7.8 & 1\\
A644 &10.310 & 0.0704 & 0.0750 & $7.7\pm0.1$ & 2054 & 25.48 & $9.4\pm0.29$ & 132.4 & 2\\
A665 & 37.066 & 0.1819 & 0.0431 & $8.0\pm0.2$ & 1987 & 10.82 & $5.6\pm0.18$ & 134.6 & 1\\
A1068 & 10.822 & 0.1375 & 0.0173 & $4.9\pm0.17$ & 1587 & 10.89 & $15.0\pm0.24$ & 9.1 & 1 \\
A1651 & 7.630 & 0.084945 & 0.0156 & $6.7\pm0.2$ & 1913 & 20.00 & $8.8\pm0.50$ & 89.5 & 2\\
A1689 & 14.291 & 0.1832 & 0.0186 & $9.2\pm0.2$ & 2126 & 11.51 & $13.8\pm0.22$ & 78.4 & 1\\
A1795 & 35.494 & 0.06248 & 0.0121 & $6.02\pm0.08$ & 1828 & 25.31 & $20.1\pm0.12$ & 19.0 & 1\\
A1991 & 21.956 & 0.0586 & 0.0248 & $2.4\pm0.1$ & 1064 & 15.64 & $16.1\pm0.22$ & 1.5 & 1\\
A2029 & 13.089 & 0.07728 & 0.0323 & $7.7\pm0.2$ & 2054 & 23.40 & $20.2\pm0.20$ & 10.5 & 1\\
A2142 & 19.410 & 0.0909 & 0.0383 & $9.0\pm0.3$ & 2209 & 21.73 & $10.3\pm0.17$ & 68.1 & 3\\
A2163 & 7.267 & 0.203 & 0.109 & $18.8\pm1.3$ & 3008 & 15.01 & $8.2\pm0.92$ & 438.0 & 2\\
A2204 & 5.346 & 0.1526 & 0.0561 & $8.3\pm0.2$ & 2057 & 12.93 & $33.3\pm0.76$ & 9.7 & 1\\
A2218 & 43.179 & 0.1756 & 0.0266 & $6.7\pm0.3$ & 1825 & 10.22 & $4.6\pm0.10$ & 288.6 & 1\\
A2255 & 13.676 & 0.0806 & 0.0250 & $6.1\pm0.1$ & 1817 & 19.9 & $2.3\pm0.32$ & 529.1 & 2\\
A2256 & 17.000 & 0.0581 & 0.0418 & $6.2\pm0.1$ & 1865 & 27.63 & $3.0\pm0.47$ & 349.6 & 1\\
A2597 & 7.426 & 0.0852 & 0.0246 & $3.64\pm0.06$ & 1405 & 14.65 & $18.0\pm0.22$ & 10.6 & 1 \\
A3112 & 7.829 & 0.07525 & 0.0137 & $4.8\pm0.1$ & 1613 & 18.82 & $18.3\pm0.26$ & 11.4 & 1\\
A3158 & 3.123 & 0.0597 & 0.0138 & $5.1\pm0.1$ & 1681 & 24.27 & $3.8\pm0.20$ & 166.0 & 1\\
A3266 & 13.967 & 0.0589 & 0.0158 & $9.2\pm0.3$ & 2260 & 33.05 & $5.3\pm0.49$ & 72.5 & 3\\
A3558 & 28.751 & 0.048 & 0.0402 & $5.06\pm0.05$ & 1687 & 29.89 & $7.2\pm0.23$ & 126.2 & 1\\
A3562 & 20.518 &  0.049 & 0.0376 & $4.8\pm0.3$ & 1635 & 28.41 & $5.7\pm0.26$ & 77.4 & 3\\
A3667 & 12.462 & 0.0556 & 0.0452 & $5.31\pm0.05$ & 1721 & 26.56 & $4.5\pm0.36$ & 160.4 & 2\\
A4059 & 5.684 & 0.0475 & 0.0122 & $4.07\pm0.08$ & 1513 & 27.08 & $4.7\pm0.33$ & 7.1 & 1\\
Hydra A & 18.541 & 0.0539 & 0.0468 & $4.0\pm0.06$ & 1495 & 23.75 & $22.1\pm0.17$ & 13.3 & 1\\
MKW 3s & 9.781 & 0.045 & 0.0272 & $3.52\pm0.06$ & 1409 & 26.54 & $13.5\pm0.22$ & 23.9 & 1\\
PKS 0745-191 & 9.627 & 0.1028 & 0.405 & $8.4\pm0.3$ & 2121 & 18.70  & $31.9\pm0.45$ & 12.4 & 1\\
Triangulum & 7.343 & 0.051 & 0.114 & $8.9\pm0.2$ & 2229 & 37.31 & $5.9\pm0.79$ & 313.0 & 1 \\
\, \\
\hline
\end{tabular}
\end{table}
\end{landscape}
\end{center}

\appendix
\section{Determination of azimuthal scatter profiles}
\label{appscat}

The azimuthal scatter \citep{vazzascat} is defined as the relative scatter in surface brightness between various sectors (see Sect. \ref{azmet}),

\begin{equation}\Sigma^2=\frac{1}{N}\sum_{i=1}^N \frac{(SB_i-\langle SB \rangle)^2}{\langle SB \rangle}.\label{stotapp}\end{equation}

\noindent In practice, computing this quantity is difficult, since the statistical fluctuations of the surface brightness introduce a contribution to the scatter that is actually dominant in the outer regions. To estimate the intrinsic level of azimuthal scatter, we used two different complementary methods, which we describe in more detail here.

\subsection{Subtraction of the statistical scatter}
\label{metone}

Since the statistical fluctuations of the data also introduce a certain level of scatter, it must be noted that the quantity computed through Eq. \ref{stotapp} gives the sum of the statistical and intrinsic scatter,

\begin{equation} \Sigma^2=\Sigma_{int}^2+\Sigma_{stat}^2.\label{statscat}\end{equation}

\noindent The statistical scatter $\Sigma_{stat}$ is given by the mean of the individual relative errors,

\begin{equation} \Sigma_{stat}^2=\frac{1}{N}\sum_{i=1}^N \frac{\sigma_i^2}{\langle SB\rangle^2},\label{sigmastat}\end{equation}

\noindent and must be subtracted from Eq. \ref{stot} to estimate the level of intrinsic scatter. The validity of Eq. \ref{sigmastat} for the statistical scatter was verified through a set of simulations of a source with no intrinsic scatter.

The uncertainties in the scatter are then estimated through Monte Carlo simulations. Namely, the surface-brightness values in the $N$ sectors are randomized, and the scatter is recomputed each time. This procedure is applied $10^3$ times, and the error on the scatter is defined as the RMS of the distribution around the mean value.

\subsection{Maximum likelihood estimation}
\label{met2}

\begin{figure}
\resizebox{\hsize}{!}{\includegraphics[angle=270]{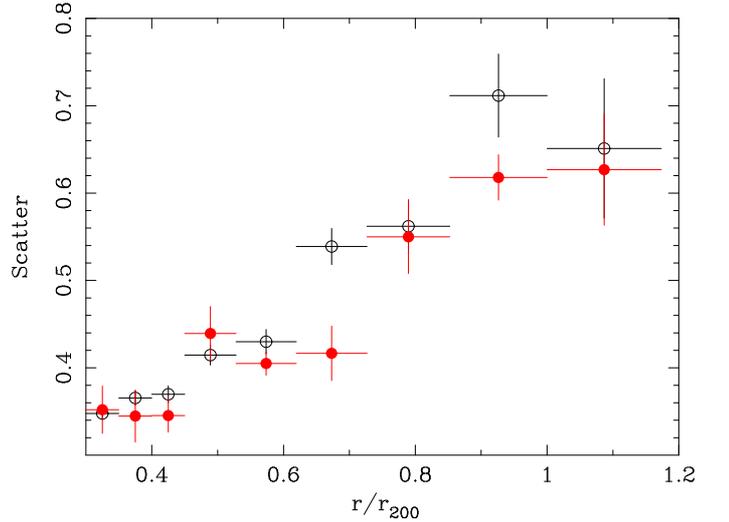}}
\caption{Comparison between the mean azimuthal scatter profiles computed using the direct method (black, see Sect. \ref{metone}) and the alternative method using a maximum likelihood estimator (red, see Sect. \ref{met2}).}
\label{compscat}
\end{figure}

To check the validity of our approach we performed an independent analysis of the scatter. We model the intrinsic scatter in the form of a Gaussian. We use a maximum likelihood algorithm \citep{maccacaro} to fit the data, where the free parameters are the mean and the intrinsic scatter (i.e. the standard deviation of the Gaussian). The methods described in both Sect. \ref{metone} and this appendix were applied to the surface brightness distribution within the annuli of each cluster (see Sect \ref{azmet} for details). Intrinsic scatter profiles from different objects were rebinned onto a common grid in units of $r_{200}$ and  
stacked. In Fig. \ref{compscat} we compare the intrinsic scatters measured with the two methods. The profiles are very similar, the general trend towards 
increasing scatter with radius is recovered with both methods. The only bin where a significant difference is observed is around $0.7r_{200}$. This comparison therefore provides a confirmation of our scatter analysis using two very different methods.

\section{Notes on individual objects}
\label{appendix}

\begin{itemize}
\item
\textbf{\textit{A85:}}\\
A subcluster located $\sim10^\prime$ south of the cluster center is currently merging with the main cluster. This substructure was masked for the analysis.

\item
\textbf{\textit{A401:}}\\
The cluster is connected through a filament to its neighbor A399, located $\sim35^{\prime}$ south-west of the center of A401. We extracted the surface-brightness profile in a sector of position angle 340-250$^{\circ}$ to avoid any contamination of A399 to our measurement of the CXB.

\item
\textbf{\textit{A478:}}\\
The combination of a favorable temperature/redshift and a good-quality \emph{ROSAT} observation allows us to reach the highest signal-to-noise ratio in the sample at $r_{200}$ for this strong CC cluster. As a result, the data from this cluster may contribute strongly when a weighted mean is performed.

\item
\textbf{\textit{A644:}}\\
This NCC cluster exhibits an unusual decreasing azimuthal scatter profile, showing large (close to 100\%) scatter in its central regions, but no significant scatter around $r_{200}$.

\item
\textbf{\textit{A2029:}}\\
A probable filament connects A2029 to A2033, located $\sim35^{\prime}$ north of the center of A2029. The surface-brightness profile was extracted in a sector with position angle 140-80$^{\circ}$ to measure the CXB level.

\item
\textbf{\textit{A2142:}}\\
Several PSPC observations of this famous cold-front cluster exist. For this work, we used the longest available observation, which was pointed $16^\prime$ south of the center of A2142. This is the only case in the sample for which the observation was not pointed on the target.

\item
\textbf{\textit{A3558 and A3562:}}\\
These two clusters are located in the Shapley supercluster and connected by a filament. Consequently, they show an unusually high azimuthal scatter in the outskirts. The CXB level was estimated by excluding the direction of the filament.

\item
\textbf{\textit{A3667:}}\\
This very disturbed cluster shows the highest emission-measure and density in the sample beyond $\sim0.2r_{200}$, and hence it could bias our average profiles, in particular when computing the difference between the CC and NCC classes. However, removing it from the sample did not lead to any significant difference, either quantitative or qualitative.

\item
\textbf{\textit{A4059:}}\\
This is the most azimuthally-symmetric cluster in the sample. The azimuthal scatter for this cluster is consistent with 0 at all radii.

\item
\textbf{\textit{Hydra A:}}\\
A tail of emission (filament?) extends out to $\sim20^{\prime}$ south-east of the cluster core. This leads to a very high azimuthal scatter ($>100\%$) around $r_{200}$.

\end{itemize}

\section{Mean emission-measure profiles}
\label{appem}

In Table \ref{emcomp} we give the mean self-similar scaled emission-measure profiles for the CC and NCC classes and the whole sample, as shown in Fig. \ref{meanem}.

\begin{table*}
\caption{\label{emcomp}Data of Fig. \ref{meanem}: mean self-similar scaled emission-measure profiles for the whole sample and for the CC and NCC classes, in units of cm$^{-6}$ Mpc}
\begin{tabular}{ccccc}
\hline
$R_{in}$ & $R_{out}$ & Total & CC & NCC\\ 
\hline
\hline
0 & 0.02 & $(1.80\pm0.01)\cdot10^{-5}$ & $(9.48\pm0.05)\cdot10^{-5}$ & $(1.13\pm0.01)\cdot10^{-5}$\\ 
0.02 & 0.04 & $(1.26\pm0.01))\cdot10^{-5}$ & $(4.83\pm0.02)\cdot10^{-5}$ & $(8.32\pm0.06)\cdot10^{-6}$\\ 
0.04 & 0.06 & $(9.63\pm0.04)\cdot10^{-6}$ & $(2.28\pm0.01)\cdot10^{-5}$ & $(6.90\pm0.04)\cdot10^{-6}$\\ 
0.06 & 0.08 & $(7.39\pm0.03)\cdot10^{-6}$ & $(1.23\pm0.01)\cdot10^{-5}$ & $(5.70\pm0.03)\cdot10^{-6}$\\ 
0.08 & 0.1 & $(5.45\pm0.02)\cdot10^{-6}$ & $(7.72\pm0.04)\cdot10^{-6}$ & $(4.49\pm0.02)\cdot10^{-6}$\\ 
0.1 & 0.12 & $(4.12\pm0.02)\cdot10^{-6}$ & $(5.27\pm0.03)\cdot10^{-6}$ & $(3.52\pm0.02)\cdot10^{-6}$\\ 
0.12 & 0.14 & $(3.20\pm1.36)\cdot10^{-6}$ & $(3.63\pm0.02)\cdot10^{-6}$ & $(2.91\pm0.02)\cdot10^{-6}$\\ 
0.14 & 0.16 & $(2.47\pm0.01)\cdot10^{-6}$ & $(2.60\pm0.02)\cdot10^{-6}$ & $(2.37\pm0.01)\cdot10^{-6}$\\ 
0.16 & 0.18 & $(1.91\pm0.01)\cdot10^{-6}$ & $(1.95\pm0.01)\cdot10^{-6}$ & $(1.88\pm0.01)\cdot10^{-6}$\\ 
0.18 & 0.2 & $(1.51\pm0.01)\cdot10^{-6}$ & $(1.48\pm0.01)\cdot10^{-6}$ & $(1.54\pm0.01)\cdot10^{-6}$\\ 
0.2 & 0.22 & $(1.23\pm0.01)\cdot10^{-6}$ & $(1.19\pm0.01)\cdot10^{-6}$ & $(1.26\pm0.01)\cdot10^{-6}$\\ 
0.22 & 0.24 & $(1.02\pm0.01)\cdot10^{-6}$ & $(9.47\pm0.09)\cdot10^{-7}$ & $(1.07\pm0.01)\cdot10^{-6}$\\ 
0.24 & 0.26 & $(8.40\pm0.05)\cdot10^{-7}$ & $(7.61\pm0.08)\cdot10^{-7}$ & $(8.95\pm0.07)\cdot10^{-7}$\\ 
0.26 & 0.29 & $(6.91\pm0.05)\cdot10^{-7}$ & $(6.09\pm0.07)\cdot10^{-7}$ & $(7.59\pm0.06)\cdot10^{-7}$\\ 
0.29 & 0.31 & $(5.32\pm0.04)\cdot10^{-7}$ & $(4.73\pm0.06)\cdot10^{-7}$ & $(5.77\pm0.05)\cdot10^{-7}$\\ 
0.31 & 0.34 & $(4.30\pm0.04)\cdot10^{-7}$ & $(3.74\pm0.06)\cdot10^{-7}$ & $(4.70\pm0.05)\cdot10^{-7}$\\ 
0.34 & 0.38 & $(3.20\pm0.03)\cdot10^{-7}$ & $(2.77\pm0.04)\cdot10^{-7}$ & $(3.60\pm0.04)\cdot10^{-7}$\\ 
0.38 & 0.41 & $(2.49\pm0.02)\cdot10^{-7}$ & $(2.10\pm0.04)\cdot10^{-7}$ & $(2.76\pm0.03)\cdot10^{-7}$\\ 
0.41 & 0.45 & $(1.86\pm0.02)\cdot10^{-7}$ & $(1.57\pm0.03)\cdot10^{-7}$ & $(2.11\pm0.03)\cdot10^{-7}$\\ 
0.45 & 0.50 & $(1.48\pm0.02)\cdot10^{-7}$ & $(1.27\pm0.03)\cdot10^{-7}$ & $(1.63\pm0.02)\cdot10^{-7}$\\ 
0.50 & 0.55 & $(1.07\pm0.02)\cdot10^{-7}$ & $(9.05\pm0.24)\cdot10^{-8}$ & $(1.18\pm0.02)\cdot10^{-7}$\\ 
0.55 & 0.60 & $(7.99\pm0.14)\cdot10^{-8}$ & $(6.82\pm0.22)\cdot10^{-8}$ & $(8.87\pm0.19)\cdot10^{-8}$\\ 
0.60 & 0.66 & $(5.73\pm0.12)\cdot10^{-8}$ & $(4.97\pm0.18)\cdot10^{-8}$ & $(6.30\pm0.16)\cdot10^{-8}$\\ 
0.66 & 0.72 & $(4.28\pm0.11)\cdot10^{-8}$ & $(3.78\pm0.17)\cdot10^{-8}$ & $(4.62\pm0.14)\cdot10^{-8}$\\ 
0.72 & 0.79 & $(3.06\pm0.11)\cdot10^{-8}$ & $(2.75\pm0.18)\cdot10^{-8}$ & $(3.21\pm0.13)\cdot10^{-8}$\\ 
0.79 & 0.87 & $(2.23\pm0.10)\cdot10^{-8}$ & $(1.77\pm0.16)\cdot10^{-8}$ & $(2.51\pm0.13)\cdot10^{-8}$\\ 
0.87 & 0.95 & $(1.35\pm0.09)\cdot10^{-8}$ & $(8.57\pm1.49)\cdot10^{-9}$ & $(1.63\pm0.11)\cdot10^{-8}$\\ 
0.95 & 1.05 & $(7.77\pm0.85)\cdot10^{-8}$ & $(5.85\pm1.40)\cdot10^{-9}$ & $(8.88\pm1.07)\cdot10^{-9}$\\ 
1.05 & 1.15 & $(5.32\pm0.80)\cdot10^{-8}$ & $(4.19\pm1.35)\cdot10^{-9}$ & $(5.92\pm0.99)\cdot10^{-9}$\\ 
1.15 & 1.26 & $(4.74\pm0.81)\cdot10^{-8}$ & $(3.75\pm1.40)\cdot10^{-9}$ & $(5.24\pm0.97)\cdot10^{-9}$\\ 
\hline
\end{tabular}

\end{table*}

\section{Computing the gas fraction from density profiles}

The gas fraction in the observations and in the simulated clusters within an overdensity $\Delta$ can be computed directly from the profiles presented in Fig. \ref{denssimnr}. Indeed, by definition,

\begin{equation}M_\Delta=\Delta\rho_{crit}\frac{4}{3}\pi r_{\Delta}^3,\end{equation}

\noindent where $\rho_{crit}=\frac{3H_0^2}{8\pi G}=9.2\times 10^{-30}$ g cm$^{-3}$. Then,

\begin{equation}f_{gas,\Delta}=\frac{M_{gas,\Delta}}{M_\Delta}=\frac{3}{\Delta\rho_{crit}r_{\Delta}^3}\int_0^{r_\Delta} \rho_{gas}(r)r^2\,dr \end{equation}

\noindent Making the substitution $x=\frac{r}{r_{\Delta}}$, we find the convenient formula

\begin{equation}f_{gas,\Delta}=\frac{3}{\Delta\rho_{crit}}\int_0^1\rho_{gas}(x)x^2\, dx.\end{equation}

\end{document}